\documentclass[a4paper,11pt]{article}
\usepackage{latexsym}
\usepackage{amsmath}
\usepackage{amssymb}
\usepackage{graphicx}
\usepackage{wrapfig}
\begin{document}

\noindent
SAGA-HE-211

\bigskip

\centerline{\bf Effective Meson Masses, Effective Meson-Nucleon Couplings}
\centerline{\bf and Neutron Star Radii}

\bigskip

\centerline{Kunito Tuchitani$^{3}$, Yasushi Horinouchi$^{1}$, Taihei Iwamitsu
$^{4}$, Ken-ichi Makino$^{1}$, }

\centerline{Nobuo Noda$^{1}$, Hiroaki Kouno$^{1}$, Akira Hasegawa$^{2}$
 and Masahiro Nakano$^5$}

\noindent
$^1${ Department of Physics, Saga University, Saga 840-8502, Japan}

\noindent
$^2${ Saga University, Saga 840-8502, Japan}

\noindent
$^3${ Saga High School of Technology, Saga 840, Japan}

\noindent
$^4${ Namura Information Systems Co., LTD, Imari-shi, Saga 848-0121, Japan}

\noindent
$^5${ University of Occupational and Environmental Health, Kitakyusyu 807, Japan}

\bigskip

\centerline{\bf Abstract}
\noindent
Using the generalized mean field theory, we have studied the relation among the effective meson masses, the effective meson-nucleon couplings and the equation of state (EOS) in asymmetric nuclear matter. 
If the effective $\omega$-meson mass becomes smaller at high density, 
the EOS becomes stiffer. 
However, if we require that the $\omega$-meson mean field is proportional to the baryon density, the effective $\omega$-nucleon coupling automatically becomes smaller at the same time as the effective $\omega$-meson mass becomes smaller. 
Consequently, the EOS becomes softer. 
A similar relation is found for the effective $\rho$-meson mass and the effective $\rho$-nucleon coupling. 
We have also studied the relation among the effective meson masses, the effective meson-nucleon couplings and a radius $R$ of a neutron star. 
The $R$ depends somewhat on the value of the effective $\omega$-meson mass and the effective $\omega$-nucleon coupling. 
The ambiguity of $R$ is a few hundred meters if $\vert {m_{\omega 0}^*}^2-m_\omega^2\vert\sim 0.1m_\omega^2$ at the normal density. 

\bigskip

%\begin{flushleft}
%%%%%%%%%%%%%%%%%%%%%%%%%%%%%%
\section{Introduction}
%%%%%%%%%%%%%%%%%%%%%%%%%%%%%%

Medium effects on the hadron masses and couplings are much interested in the hadron and nuclear physics. \cite{rf:Brown,rf:Hatsuda,rf:Hatsuda2} 
In particular, the vector meson mass reduction attracts a great deal of attention because they are related to the chiral symmetry restoration. 
Because of its short life time, the reduction of the $\rho$-meson mass is expected to be a signal of the hot and dense matter which may be produced in the high-energy heavy ion collisions. \cite{rf:Tserruya}  

On the other hand, the $\omega$-meson is important for the nuclear structure. 
The results of the relativistic Brueckner Hartree Fock calculation \cite{rf:Brockmann,rf:deJong} are well described by the $\omega$-meson self-interaction \cite{rf:Bodmer,rf:Sugahara} or the decrease of the $\omega$-nucleon couplings in medium. \cite{rf:Brockmann2} 
In the paper \cite{rf:Tuchitani}, Tuchitani et al. showed analytically that the enhancement of the effective $\omega$-meson mass caused by the $\omega$-meson self-interaction softens the equation of state (EOS). 
They also showed that the reduction of the effective $\omega$-nucleon coupling is related with the enhancement of the effective $\omega$-meson mass if it is caused by the effective multi $\omega$-nucleon interaction. 
However, it seems that such an enhancement of the effective $\omega$-meson mass is not consistent with the chiral symmetry restoration in medium. 
It is reported that the reduction of the effective vector meson mass makes the EOS stiffer. \cite{rf:Weber,rf:Hyun} 
Recently, Kouno et al. showed that the effective $\omega$-nucleon coupling automatically decreases as the effective $\omega$-meson mass decreases, if we require that the $\omega$-meson mean field is proportional to the baryon density. \cite{rf:Kouno} 
In that case, the EOS becomes softer in spite of the reduction of the effective $\omega$-meson mass. 

On the other hand, the EOS also becomes softer, if there is a exotic matter such as pion condensation \cite{rf:Nakano} or quark matters. \cite{rf:Alford} 
If an observable such as a neutron star radius indicates that the EOS is too soft to be described by the hadron EOS, it is a proof of the existence of such an exotic matter. 
In this meaning, it is important to examine how soft the "normal" hadon EOS can be without such an exotic matter. 

In this paper, we study the EOS of asymmetric nuclear matter using the generalized mean field theory proposed by Tuchitani et al. \cite{rf:Tuchitani} and study the relation among the effective meson masses, the effective meson-nucleon couplings and the EOS. 
If the effective $\omega$-meson mass becomes smaller (larger) at high density, 
the EOS becomes stiffer (softer). 
However, as is in the case of the symmetric nuclear matter, \cite{rf:Kouno} if we require that the $\omega$-meson mean field is proportional to the baryon density, the effective $\omega$-nucleon coupling automatically becomes smaller (larger) at the same time as the effective $\omega$-meson mass becomes smaller (larger). 
The effect of the reduction (enhancement) of the effective $\omega$-nucleon coupling overcomes the effect of the reduction (enhancement) of the effective $\omega$-meson mass. 
Consequently, the EOS becomes softer (stiffer). 
A similar relation is found for the effective $\rho$-meson mass and the effective $\rho$-nucleon coupling. 

We also study the relation among the effective meson masses, the effective meson-nucleon couplings and a radius of a neutron star. 
We found that the radius of the neutron star depends somewhat on the value of the effective $\omega$-meson mass and the effective $\omega$-nucleon coupling. 

This paper is organized as follows. 
In Sec. 2, we formulate the generalized mean field theory in the asymmetric nuclear matter. 
In Sec. 3, we show that the analytical representations for the incompressibility and the symmetry energy for nuclear matter. 
In Sec. 4, we numerically analyze the relation between the effective meson masses, the effective meson-nucleon couplings, the EOS and a radius of a neutron star. 
Sec. 5 is devoted to a summary. 

%%%%%%%%%%%%%%%%%%%%%%%%%%%%%%%
\section{Effective Lagrangian and Equation of Motion}
%%%%%%%%%%%%%%%%%%%%%%%%%%%%%%%

We start with the following Lagrangian of $\sigma$-$\omega$-$\rho$ model. 
%%%%%%%%
\begin{eqnarray}
L&=&\bar{\psi}\left[\gamma^\mu\{i\partial_\mu+\Sigma_\mu (\sigma, \omega, \rho )+\tau^a\Sigma^{a}_\mu(\sigma,\omega,\rho)\} -\{m+\Sigma_{\rm s}(\sigma ,\omega, \rho)\}\right]\psi
\nonumber\\
&+&{1\over{2}}\partial^\mu\sigma\partial_\mu\sigma
-{1\over{4}}F_{\mu\nu}F^{\mu\nu}-{1\over{4}}G_{\mu\nu}^aG^{\mu\nu a}
\nonumber\\
&-&U_M(\sigma,\omega,\rho );
\nonumber\\
&F_{\mu\nu}&=\partial_\mu\omega_\nu -\partial_\nu\omega_\mu,~~~~~
G_{\mu\nu}^a=\partial_\mu \rho_\nu^a-\partial_\nu \rho_\nu^a-g_\rho\epsilon^{abc}\rho_\mu^b\rho_\nu^c, 
\label{eq:E1}
\end{eqnarray}
%%%%%%%%
where $\psi$, $\sigma$, $\omega_\mu$ $\rho_\mu^a$, $m$ and $\tau^a~(a=1,2,3)$ are the nucleon field, the $\sigma$-meson field, the $\omega$-meson field, the $\rho$-meson field, the nucleon mass and the Pauli matries in the isospin space, respectively. 
The $\Sigma_{\rm s}$, $\Sigma_\mu$, $\Sigma^a_\mu$ and $U_{\rm M}$ are functions of $\sigma$, $\omega^\mu$ and $\rho^a_\mu$.  
The $\Sigma_{\rm s}$, $\Sigma_\mu$ and $\Sigma_\mu^a$ are the nucleon self-energies, while the $U_{\rm M}$ is the mesonic potential which includes the meson mass terms, the term of the $\sigma$-meson self-interaction, the term of the $\omega$-meson self-interaction and so forth. 

We regard the Lagrangian (\ref{eq:E1}) as an effective one in which the quantum effects of the vacuum fluctuations have been already included. 
Therefore, in principle, there are a limitless number of parameters, namely, effective couplings in $\Sigma_{\rm s}$, $\Sigma_\mu$, $\Sigma_\mu^a$ and $U_{M}$. 

We remark that the effective Lagrangian (\ref{eq:E1}) includes the large classes of the relativistic nuclear models. 
It includes the original Walecka model,\cite{rf:Walecka,rf:Serot} the relativistic Hartree approximation,\cite{rf:Chin,rf:Serot} the nonlinear $\sigma$-$\omega$ model with $\sigma$-meson self-interactions \cite{rf:Boguta,rf:Reinhard,rf:Waldhauser,rf:Sharma,rf:Kouno1,rf:Lalazissis,rf:Iwasaki} and the $\omega$-meson self-interaction, \cite{rf:Bodmer,rf:Sugahara,rf:Kouno2} the model including $\sigma$-$\omega$ meson interaction \cite{rf:Moncada}, the model with the $\sigma$-$\rho$ and the $\omega$-$\rho$ interactions and so forth. 
It also includes the Zimanyi and Mozkowski (ZM) model after the fermion wave function is rescaled. \cite{rf:Zimanyi} 
In the quark-meson coupling model (QMC), \cite{rf:Guichon,rf:Saito,rf:Saito2}, the nucleon self-energy is calculated using the bag model. \cite{rf:MIT} 

Starting from the effective Lagrangian (\ref{eq:E1}), we calculate the density effects in nuclear matter. 
We use the mean field approximation. 
In the uniform and rotational invariant nuclear matter, the ground-state expectation value of the spatial component of the $\omega$-meson fields is zero. 
Therefore, below, we only work with $<\omega^0>$, the expectation value of the time-like component of the $\omega$-meson field, and write it in the symbol of $\omega$. 
Similarly, for the $\rho$-meson field, only the time-like neutral component $<\rho_0^3>$ does not vanish. 
We write it in the symbol of $\rho$. 
We also write the expectation value $<\sigma >$ in the symbol of $\sigma$.  

The $\Sigma_{\rm s}$, $\gamma^\mu\Sigma_\mu$ and $\gamma^\mu\tau^a\Sigma_\mu^a$ are the self-energies of the nucleon. (See Fig. 1(a).) 
Since $\Sigma_i(i=1,2,3)$ and $\Sigma_i^a(i=1,2,3)$ has at least one spatial component of the vector meson fields, they also become zero in the mean field approximation. 
Below, we write $\Sigma_0$ as $\Sigma_{\rm v}$. 
Similarly, in $\Sigma^a_\mu$, only the time-like neutral component $\Sigma_0^3$ does not vanish. 
Below, we write it in the symbol of $\Sigma_{\rm r}$. 
In the Lagrangian (\ref{eq:E1}), we have neglected the other parts of the self-energies which vanish in the mean-field approximation. 
(For example, the tensor part $\bar{\psi}[\gamma^\mu,\gamma^\nu ]\Sigma_{\mu\nu}\psi$ vanishes in the mean field approximation, since $\Sigma_{\mu\nu}$ is anti-symmetric in the subscripts $\mu$ and $\nu$ and includes at least one $\omega^i$. )

In the mean field approximation, the nucleon (N) propagator $G^{\rm N}(k)$ (N=p or n) is given by \cite{rf:Walecka, rf:Chin, rf:Serot}
%%%%%%%%
\begin{equation}
G_{\rm N}(k)=G_{\rm F}^{\rm N}(k)+G_{\rm N}^{\rm D}(k)
\label{eq:E1a}
\end{equation}
%%%%%%%%
with the Feynman part 
%%%%%%%%
\begin{equation}
G^{\rm F}_{\rm N}(k)=(\gamma^\mu k_\mu^*+m^*){-1\over{-{k^*}^2+{m^*}^2-i\epsilon}}
\label{eq:E1af}
\end{equation}
%%%%%%%%
and the density part 
%%%%%%%%
\begin{equation}
G^{\rm D}_{\rm N}(k)=(\gamma^\mu k_\mu^*+m^*){i\pi\over{E_k^*}}\delta ({k^*}^0-E_k^*)\theta (k_{\rm FN}-\vert {\bf k}\vert ), 
\label{eq:E1ad}
\end{equation}
%%%%%%%%
where $m^*=m+\Sigma_{\rm s}$, $E_k^*=\sqrt{{\bf k}^2+{m^*}^2}$, ${k^*}^\mu=(k^0+\Sigma_{\rm v}+\Sigma_{\rm r},{\bf k})$ for proton, ${k^*}^\mu=(k^0+\Sigma_{\rm v}-\Sigma_{\rm r},{\bf k})$ for neutron, respectively, and $k_{{\rm FN}}$ is the Fermi momentum. 
Since the effects of vacuum fluctuations have been already included in the effective Lagrangian (\ref{eq:E1}), we use only the density part $G^{\rm D}_{\rm N}(k)$ to evaluate the density effects. 

Using the propagator $G^{\rm D}_{\rm N}(k)$, we get 
the baryon density 
\begin{equation}
\rho_{\rm B}=\rho_{\rm p}+\rho_{\rm n}, 
\label{eq:E2a}
\end{equation}
where $\rho_{\rm p}$ and $\rho_{\rm n}$ are the proton and neutron densities, respectively. 
The $\rho_{\rm N}$ (N=p or n) is given by 
\begin{eqnarray}
\rho_N 
&=& <\bar{\psi}\gamma^0\psi >_N
=-i\int{d^4k\over{(2\pi)^4}}{\rm Tr}[\gamma^0G^{\rm D}_{\rm N}(k)]
={1\over{3\pi^2}}k_{{\rm FN}}^3. 
\label{eq:3}
\end{eqnarray}

The scalar density is also given by 
\begin{equation}
\rho_{\rm s}=\rho_{\rm sp}+\rho_{\rm sn}, 
\label{eq:3a}
\end{equation}
where $\rho_{\rm sp}$ and $\rho_{\rm sn}$ are the proton and the neutron scalar densities, respectively. 
The $\rho_{\rm sN}$ (N=p or n) is given by 
\begin{eqnarray}
\rho_{\rm sN}(\rho_{\rm N},m^*)&=&<\bar{\psi}\psi >_N
=-i\int{d^4k\over{(2\pi)^4}}{\rm Tr}[G^{\rm D}_N(k)]
\nonumber\\
&=&{1\over{2\pi^2}}m^*\left\{k_{\rm FN}E_{\rm FN}^*-{m^*}^2\ln{\left({k_{\rm FN}+E_{\rm FN}^*\over{m^*}}\right)}\right\}, 
\label{eq:4}
\end{eqnarray}
%%%%%%%%
where $E_{\rm FN}^*=\sqrt{k_{\rm FN}^2+{m^*}^2}$. 
The energy density of the nuclear matter is given by 
\begin{eqnarray}
\epsilon&=&\epsilon_{\rm p+n}(\rho_{\rm B},\rho_3,\sigma,\omega,\rho )+U_{\rm M}(\sigma,\omega,\rho )
-\Sigma_{\rm v}(\sigma,\omega,\rho )\rho_{\rm B}-\Sigma_{\rm r}(\sigma,\omega,\rho)\rho_3
\nonumber\\
&=&\epsilon_{\rm p}(\rho_p,m^*)+\epsilon_{\rm n}(\rho_n,m^*)+U_{\rm M}(\sigma,\omega,\rho )
-\Sigma_{\rm v}(\sigma,\omega,\rho )\rho_{\rm B}-\Sigma_{\rm r}(\sigma,\omega,\rho)\rho_3, 
\nonumber\\
\label{eq:5}
\end{eqnarray}
where 
\begin{eqnarray}
\epsilon_N(\rho_{\rm N},m^*)
&=&{1\over{12\pi^2}}\left\{ E_{\rm FN}^* k_{\rm FN}(3k_{\rm FN}^2+{3\over{2}}{m^*}^2)-{3\over{2}}{m^*}^4\log{\left({E_{\rm FN}^*+k_{\rm FN}\over{m^*}}\right)}\right\} 
\nonumber\\
&&\label{eq:6}
\end{eqnarray}
and the isovector density is defined as $\rho_3=\rho_{\rm p}-\rho_{\rm n}$. 
The pressure of the nuclear matter is also given by 
\begin{equation}
P=P_{\rm p}(\rho_{\rm p},m^*)+P_{\rm n}(\rho_{\rm n},m^*)-U_{\rm M}(\sigma,\omega,\rho ), 
\label{eq:7}
\end{equation}
where 
\begin{eqnarray}
P_N(\rho_{\rm N},m^*)&=&{1\over{12\pi^2}}\left\{ E_{\rm FN}^* k_{\rm FN}(k_{\rm FN}^2-{3\over{2}}{m^*}^2)+{3\over{2}}{m^*}^4\log{\left({E_{\rm FN}^*+k_{\rm FN}\over{m^*}}\right)}\right\}. 
\nonumber\\
&&\label{eq:8}
\end{eqnarray}

The chemical potentials for proton and for neutron are given by 
%%%%%%%%
\begin{equation}
\mu_{\rm p}=E_{\rm Fp}^*-\Sigma_{\rm v}(\sigma,\omega,\rho)-\Sigma_{\rm r}(\sigma,\omega,\rho )
\label{eq:9a}
\end{equation}
%%%%%%%%
and 
%%%%%%%%
\begin{equation}
\mu_{\rm n}=E_{\rm Fn}^*-\Sigma_{\rm v}(\sigma,\omega,\rho)+\Sigma_{\rm r}(\sigma,\omega,\rho ).  
\label{eq:9b}
\end{equation}
%%%%%%%%

The equation of motion for $\sigma$-meson is given by 
\begin{equation}
{\partial \epsilon(\rho_{\rm B},\rho_3,\sigma,\omega,\rho )\over{\partial \sigma}}=0. 
\label{eq:E10}
\end{equation}
Putting (\ref{eq:5}) into (\ref{eq:E10}), we get 
\begin{eqnarray}
&&{\partial \epsilon_{\rm p+n}(\rho_{\rm B},\rho_3, m^*(\sigma,\omega,\rho ))\over{\partial \sigma}}
-{\partial \Sigma_{\rm v}(\sigma,\omega,\rho )\over{\partial \sigma}}\rho_{\rm B}
\nonumber\\
&&-{\partial \Sigma_{\rm r}(\sigma,\omega,\rho )\over{\partial \sigma}}\rho_3
+{\partial U_{\rm M}(\sigma,\omega,\rho )\over{\partial \sigma}}
\nonumber\\
&=&
{\partial m^*(\sigma,\omega,\rho )\over{\partial \sigma}}
{\partial \epsilon_{\rm p+n}(\rho_{\rm B},\rho_3,m^*)\over{\partial m^*}}
-{\partial \Sigma_{\rm v}(\sigma,\omega,\rho )\over{\partial \sigma}}\rho_{\rm B}
\nonumber\\
&&-{\partial \Sigma_{\rm r}(\sigma,\omega,\rho )\over{\partial \sigma}}\rho_3
+{\partial U_{\rm M}(\sigma,\omega,\rho )\over{\partial \sigma}}
\nonumber\\
&=&-g_{\rm s\sigma}^*\rho_{\rm s}+g_{\rm v\sigma}^*\rho_{\rm B}+g_{\rm r \sigma}^*\rho_3 
+{\partial U_{\rm M}(\sigma,\omega,\rho )\over{\partial \sigma}}=0,
\label{eq:E11}
\end{eqnarray}
where we have used the relation 
%%%%%%%%
\begin{equation}
{\partial \epsilon_{\rm p+n}(\rho_{\rm B},\rho_3,m^*)\over{\partial m^*}}=\rho_{\rm s}
\label{eq:E11a}
\end{equation}
%%%%%%%%
and have defined the effective couplings for the three-point meson-nucleon interaction as 
%%%%%%%%%
\begin{equation}
g_{\rm s\sigma}^*\equiv 
-{\partial \Sigma_{\rm s}(\sigma,\omega,\rho )\over{\partial \sigma}}
,~~~
g_{\rm v\sigma}^*\equiv 
-{\partial \Sigma_{\rm v}(\sigma,\omega,\rho )\over{\partial \sigma}}
~~~{\rm and}~~~
g_{\rm r\sigma}^*\equiv 
-{\partial \Sigma_{\rm r }(\sigma,\omega,\rho )\over{\partial \sigma}}. 
\label{eq:E13}
\end{equation}
%%%%%%%%%
Note that the differentiating the nucleon self-energies with respect to the meson-field expectation value yields the effective couplings of the meson-nucleon interaction. 
One external line of the meson can be attached at the point where one meson mean field have been removed by the differentiation. 
(See Figs. 1(a) and (b).) 
In general, the effective action is a generating functional of one-particle-irreducible correlation functions, namely, effective masses and effective couplings. \cite{rf:Peskin} 

Similarly, we obtain the equation of motion for $\omega$-meson 
%%%%%%%%%
\begin{eqnarray}
-g_{\rm s\omega}^*\rho_{\rm s}+g_{\rm v\omega}^*\rho_{\rm B}+g_{\rm r \omega}^*\rho_3 
+{\partial U_{\rm M}(\sigma,\omega,\rho )\over{\partial \omega}}=0, 
\label{eq:E15}
\end{eqnarray}
%%%%%%%%%%
and the equation of motion for $\rho$-meson is given by 
%%%%%%%%%%
\begin{eqnarray}
-g_{\rm s\rho}^*\rho_{\rm s}+g_{\rm v\rho}^*\rho_{\rm B}+g_{\rm r \rho}^*\rho_3 
+{\partial U_{\rm M}(\sigma,\omega,\rho )\over{\partial \rho}}=0, \label{eq:E18}
\end{eqnarray}
%%%%%%%%%%
where we have also defined the effective couplings 
%%%%%%%%%
\begin{equation}
g_{\rm s\omega}^*\equiv 
-{\partial \Sigma_{\rm s}(\sigma,\omega,\rho )\over{\partial \omega}}
,~~~
g_{\rm v\omega}^*\equiv 
-{\partial \Sigma_{\rm v}(\sigma,\omega,\rho )\over{\partial \omega}}
~~~{\rm and}~~~
g_{\rm r\omega}^*\equiv 
-{\partial \Sigma_{\rm r }(\sigma,\omega,\rho )\over{\partial \omega}}. 
\label{eq:E16}
\end{equation}
%%%%%%%%%
and 
%%%%%%%%%
\begin{equation}
g_{\rm s\rho}^*\equiv 
-{\partial \Sigma_{\rm s}(\sigma,\omega,\rho )\over{\partial \rho}}
,~~~
g_{\rm v\rho}^*\equiv 
-{\partial \Sigma_{\rm v}(\sigma,\omega,\rho )\over{\partial \rho}}
~~~{\rm and}~~~
g_{\rm r\rho}^*\equiv 
-{\partial \Sigma_{\rm r }(\sigma,\omega,\rho )\over{\partial \rho}}. 
\label{eq:E19}
\end{equation}
%%%%%%%%%

The diagrammatic description for the fourth lines of Eqs. (\ref{eq:E11}), (\ref{eq:E15}) and (\ref{eq:E18}) is shown in Fig. 1(c). 
Although there are a limitless number of parameters in $\Sigma_{\rm s}$, $\Sigma_{\rm v}$ and $\Sigma_{\rm r}$, only nine effective couplings appear for the meson-nucleon interactions in Eqs. (\ref{eq:E11}),(\ref{eq:E15}) and (\ref{eq:E18}).  
If we put $g_{\rm v\sigma}^*=g_{\rm r\sigma}^*
=g_{\rm s\omega}^*=g_{\rm r\omega}^*=g_{\rm s\rho }^*=g_{\rm v\rho }^*=0$ and approximate $g_{\rm s\sigma}^*$, $g_{\rm v\omega}^*$ and $g_{\rm r\rho}^*$ as constants which are determined at the normal density, we have familiar equations of motion which are used in the original Walecka model, \cite{rf:Walecka,rf:Serot} the RHA calculation \cite{rf:Chin,rf:Serot} and the nonlinear $\sigma$-$\omega$(-$\rho$) model. \cite{rf:Boguta,rf:Reinhard,rf:Waldhauser,rf:Bodmer,rf:Sharma,rf:Sugahara,rf:Moncada,rf:Kouno1,rf:Kouno2,rf:Lalazissis,rf:Iwasaki}

Since the effective potential is a generating function of one-particle-irreducible correlation functions with vanishing external momentum,\cite{rf:Peskin} 
the second derivatives with respect to the meson mean fields yield the square of the effective meson masses. 
From the energy density (\ref{eq:5}), this can be directly shown as follows. 
(For the derivations, see \cite{rf:Tuchitani}.) 
%%%%%%%%
\begin{eqnarray}
{\partial^2 \epsilon (\rho_{\rm B},\rho_3,\sigma,\omega,\rho )\over{\partial \phi_i\partial \phi_j}}
&= &g_{\rm s{\phi_i}}^*g_{\rm s{\phi_j}}^*\Pi-g_{\rm s{\phi_i}{\phi_j}}^*\rho_{\rm s}
+g_{\rm v{\phi_i}{\phi_j}}^*\rho_{\rm B}+g_{\rm r{\phi_i}{\phi_j}}^*\rho_3
\nonumber\\
&+&{\partial^2 U_{\rm M}(\sigma,\omega,\rho )\over{\partial\phi_i\partial \phi_j}}, ~~~~~(i,j=1,2,3)
\label{eq:E32}
\end{eqnarray}
%%%%%%%%%
where $\phi_1=\sigma$, $\phi_2=\omega$, $\phi_3=\rho$, 
%%%%%%%%%
\begin{eqnarray}
\Pi (k_{\rm FN},\phi_i)&\equiv &-i\int{d^4k\over{(2\pi)^4}}
{\rm Tr}[G^{\rm F}_{\rm p}(k)G^{\rm D}_{\rm p}(k)+G^{\rm D}_{\rm p}(k)G^{\rm F}_{\rm p}(k)+G^{\rm D}_{\rm p}(k)G^{\rm D}_{\rm p}(k)]
\nonumber\\
&&-i\int{d^4k\over{(2\pi)^4}}
{\rm Tr}[G^{\rm F}_{\rm n}(k)G^{\rm D}_{\rm n}(k)+G^{\rm D}_{\rm n}(k)G^{\rm F}_{\rm n}(k)+G^{\rm D}_{\rm n}(k)G^{\rm D}_{\rm n}(k)]
\nonumber\\
\label{eq:E32z}
\end{eqnarray}
%%%%%%%%%
and the effective coupling $g_{\rm f\phi_i\phi_j}^*$ (f$=$s,v,r) is defined by
%%%%%%%%%
\begin{equation}
g_{\rm f{\phi_i}{\phi_j}}^*\equiv 
-{\partial^2 \Sigma_{\rm f}(\sigma,\omega,\rho )\over{\partial\phi_i\partial\phi_j}}. 
\label{eq:E33}
\end{equation}
%%%%%%%%%
The diagrammatic descriptions for Eq. (\ref{eq:E32}) are shown in  Fig. 1(d). 
If the mixing parts such as ${\partial^2 \epsilon\over{\partial \sigma\partial\omega}}$ vanish, ${\partial^2 \epsilon\over{\partial \sigma^2}}$, ${\partial^2 \epsilon\over{\partial \omega^2}}$ and ${\partial^2 \epsilon\over{\partial \rho^2}}$are the square of the effective meson masses which are defined at the zero external momentum. 

%%%%%%%%%%%%%%%%%%%%%%%%%%%%%%
\section{Incompressibility and Symmetry Energy}
%%%%%%%%%%%%%%%%%%%%%%%%%%%%%%

In this section, we derive the simple relation among the effective masses, effective couplings, the incompressibility and the symmetry energy. 

At first we study the density evolution of the expectation values $\sigma$, $\omega$ and $\rho$. 
Fixing $\rho_3$ and differentiating the equations of motion 
(\ref{eq:E11}), (\ref{eq:E15}) and (\ref{eq:E18}) with respect to $\rho_{\rm B}$, we obtain the following equation.  
%%%%%%%%
\begin{eqnarray}
&&{\partial\over{\partial \rho_{\rm B}}}{\partial \epsilon (\rho_{\rm B},\rho_3, \sigma,\omega,\rho )\over{\partial \phi_i}}
+\sum_{j=1}^3{d\phi_j\over{d\rho_{\rm B}}}{\partial^2\epsilon (\rho_{\rm B},\rho_3,\sigma,\omega,\rho )\over{\partial \phi_i\partial \phi_j}}=0
\label{eq:E34}
\end{eqnarray}
%%%%%%%%
We also get 
%%%%%%%%
\begin{eqnarray}
{\partial\over{\partial \rho_{\rm B}}}{\partial \epsilon (\rho_{\rm B},\rho_3,\sigma,\omega,\rho )\over{\partial\phi_i }}
&=&
{\partial\over{\partial \rho_{\rm B}}}{\partial \epsilon_{\rm p+n} (\rho_{\rm B}, \rho_3,m^*(\sigma,\omega,\rho ))\over{\partial\phi_i}}
\nonumber\\
&-&
{\partial\over{\partial \rho_{\rm B}}}\left({\partial \Sigma_{\rm v}(\sigma,\omega,\rho )\over{\partial \phi_i}}\rho_{\rm B}\right)
-
{\partial\over{\partial \rho_{\rm B}}}\left({\partial \Sigma_{\rm r}(\sigma,\omega,\rho )\over{\partial \phi_i}}\rho_3\right)
\nonumber\\
&=&
{\partial\over{\partial \rho_{\rm B}}}\left({\partial \Sigma_{\rm s}(\sigma,\omega,\rho )\over{\partial \phi_i}}{\partial \epsilon_{\rm p+n} (\rho_{\rm B},\rho_3, m^*)\over{\partial m^* }}\right)
\nonumber\\
&-&{\partial \Sigma_{\rm v}(\sigma,\omega,\rho )\over{\partial \phi_i}}
\nonumber\\
&=&
-g_{{\rm s}{\phi_i}}^*{\partial \rho_{\rm s}(\rho_{\rm B}, \rho_3, m^*)\over{\partial \rho_{\rm B}}}+g_{{\rm v}{\phi_i}}^* =-g_{{\rm s}{\phi_i}}^*{\hat{\gamma}}^{-1}+g_{{\rm v}{\phi_i}}^*, 
\nonumber\\
&&\label{eq:E37}
\end{eqnarray}
%%%%%%%%
where ${\hat{\gamma}}$ is a averaged effective gamma factor which is defined as 
%%%%%%%%
\begin{equation}
{\hat{\gamma}}^{-1}\equiv{\partial \rho_{\rm s}(\rho_{\rm B},\rho_3, m^*)\over{\partial \rho_{\rm B}}}
={1\over{2}}\left({m^*\over{E_{\rm Fp}^*}}+{m^*\over{E_{\rm Fn}^*}}\right). 
\label{eq:E38}
\end{equation}
%%%%%%%%
Using Eqs. (\ref{eq:E34}) and (\ref{eq:E37}),  
we obtain 
%%%%%%%%
\begin{equation}
{M^*}^2\left.{d{\bf \Phi}\over{d\rho_{\rm B}}}\right|_{\rho_3~{\rm fixed}} =-{\hat{\bf g}}, \label{eq:E41}
\end{equation}
%%%%%%%%
where 
%%%%%%%%
\begin{equation}
{\hat{\bf g}}\equiv -{\hat{\gamma}}^{-1}{\bf g}_{\rm s}^*+{\bf g}_{\rm v}^*
;~~~~~ 
{\bf g}_{\rm s}^*\equiv 
\left[
\begin{array}{c}
g_{\rm s\sigma}^* \\
g_{\rm s\omega}^* \\
g_{\rm s\rho}^*
\end{array}
\right], 
~~~~~
{\bf g}_{\rm v}^*\equiv 
\left[
\begin{array}{c}
g_{\rm v\sigma}^* \\
g_{\rm v\omega}^* \\
g_{\rm v\rho}^*
\end{array}
\right],
\label{eq:E42}
\end{equation}
%%%%%%%%
%%%%%%%%
\begin{equation}
{\bf \Phi}
\equiv 
\left[
\begin{array}{c}
\sigma \\
\omega \\
\rho
\end{array}
\right], 
\label{eq:E43}
\end{equation}
%%%%%%%%
and 
%%%%%%%%
\begin{equation}
{M^*}^2
\equiv \left[
\begin{array}{ccc}
{m_{\sigma}^*}^2 & {m_{\sigma\omega}^*}^2 & {m_{\sigma\rho}^*}^2 \\
{m_{\sigma\omega}^*}^2 & -{m_{\omega}^*}^2 & {m_{\omega\rho}^*}^2 \\
{m_{\sigma\rho}^*}^2 & {m_{\omega\rho}^*}^2 & -{m_{\rho}^*}^2
\end{array}
\right]
\equiv 
\left[
\begin{array}{ccc}
{\partial^2 \epsilon\over{\partial \sigma^2}} & {\partial^2 \epsilon\over{\partial \sigma \partial\omega}} & {\partial^2 \epsilon\over{\partial\sigma \partial\rho}} \\
{\partial^2 \epsilon\over{\partial\sigma\partial\omega}} & {\partial^2 \epsilon\over{\partial \omega^2}} & {\partial^2 \epsilon\over{\partial \omega \partial \rho}} \\
{\partial^2 \epsilon\over{\partial\sigma \partial\rho}} & {\partial^2 \epsilon\over{\partial\omega \partial\rho}} & {\partial^2 \epsilon\over{\partial \rho^2}}
\end{array}
\right] .
\label{eq:E44}
\end{equation}
%%%%%%%%
If $\det{{M^*}^2}\neq 0$, 
Eq. (\ref{eq:E41}) can be transformed as 
%%%%%%%%
\begin{equation}
\left.{d{\bf \Phi}\over{d\rho_{\rm B}}}\right|_{\rho_3~{\rm fixed}} =-({M^*}^2)^{-1}{\hat{\bf g}}. 
\label{eq:E45}
\end{equation}
%%%%%%%%
Similarly, fixing $\rho_{\rm B}$ and differentiating the equations of motion for mesons, we obtain 
%%%%%%%%
\begin{equation}
\left.{M^*}^2{d\Phi\over{d\rho_3}}\right|_{\rho_{\rm B}~{\rm fixed}}=-{\tilde{\bf g}}, \label{eq:E56}
\end{equation}
%%%%%%%%
where 
%%%%%%%%
\begin{equation}
{\tilde{\bf g}}\equiv -{\tilde{\gamma}}^{-1}{\bf g}_{\rm s}^*+{\bf g}_{\rm r}^*;~~~~~~~~
{\bf g}_{\rm r}^*\equiv 
\left[
\begin{array}{c}
g_{\rm r\sigma}^* \\
g_{\rm r\omega}^* \\
g_{\rm r\rho}^*
\end{array}
\right],~~~~~
\tilde{\gamma}^{-1}\equiv {1\over{2}}\left({m^*\over{E_{\rm Fp}^*}}-{m^*\over{E_{\rm Fn}^*}}\right). 
\label{eq:E57}
\end{equation}
%%%%%%%%
If $\det{{M^*}^2}\neq 0$, 
Eq. (\ref{eq:E56}) can be transformed as 
%%%%%%%5
\begin{equation}
\left.{d{\bf \Phi}\over{d\rho_3}}\right|_{\rho_{\rm B}~{\rm fixed}}=-({M^*}^2)^{-1}{\tilde{\bf g}}. 
\label{eq:E59}
\end{equation}
%%%%%%%%

At $\rho_3=0$, the system is symmetric for the proton and neutron and we can define the following quantities. 
%%%%%%%%
\begin{equation}
k_{\rm F}\equiv k_{\rm Fp}=k_{\rm Fn},~~~~~\mu \equiv\mu_{\rm p}=\mu_{\rm n},~~~~~E_{\rm F}^*\equiv E_{\rm Fp}=E_{\rm Fn}. 
\label{eq:63}
\end{equation}
%%%%%%%%
Using these quantities, the incompressibility $K$ is defined by  
%%%%%%%%
\begin{equation}
K\equiv 
9\rho_{\rm B0}^2\left.{d^2 (\epsilon /\rho_{\rm B} ) \over{d \rho_{\rm B}^2}}\right|_{\scriptstyle\rho_{\rm B} =\rho_{\rm B0}\atop\scriptstyle \rho_3=0}
=
9\left.{d P\over{d \rho_{\rm B}}}\right|_{\scriptstyle\rho_{\rm B} =\rho_{\rm B0}\atop\scriptstyle \rho_3=0}
=
9\rho_{\rm B0}\left.{d \mu\over{d \rho_{\rm B}}}\right|_{\scriptstyle\rho_{\rm B} =\rho_{\rm B0}\atop\scriptstyle \rho_3=0}, 
\label{eq:E64}
\end{equation}
%%%%%%%%
where $\rho_{\rm B0}$ is the normal baryon density. 
Using 
$\mu =E_{\rm F}^*-\Sigma_{\rm v}=\sqrt{k_{\rm F}^2+{m^*}^2}-\Sigma_{\rm v}$ and Eq. (\ref{eq:E45}),  
we get 
%%%%%%%%
\begin{eqnarray}
{K(\rho_{\rm B})\over{9\rho_{\rm B}}}&\equiv& \left.{1\over{\rho_{\rm B}}}{dP\over{d\rho_{\rm B}}}\right|_{\rho_3=0}={d\mu\over{d\rho_{\rm B}}}
=
{dk_{\rm F}\over{d\rho}}
{k_{\rm F}\over{E_{\rm F}^*}}
\nonumber\\
&+&{m^*\over{E_{\rm F}^*}}\left(
 {\partial \Sigma_{\rm s}(\sigma,\omega,\rho )\over{\partial \sigma}}{d\sigma\over{d\rho_{\rm B}}}
+{\partial \Sigma_{\rm s}(\sigma,\omega,\rho )\over{\partial \omega}}{d\omega\over{d\rho_{\rm B}}}
+{\partial \Sigma_{\rm s}(\sigma,\omega,\rho )\over{\partial \rho}}{d\rho\over{d\rho_{\rm B}}}\right)
\nonumber\\
&-&\left(
 {\partial \Sigma_{\rm v}(\sigma,\omega,\rho )\over{\partial \sigma}}{d\sigma\over{d\rho_{\rm B}}}
+{\partial \Sigma_{\rm v}(\sigma,\omega,\rho )\over{\partial \omega}}{d\omega\over{d\rho_{\rm B}}}
+{\partial \Sigma_{\rm v}(\sigma,\omega,\rho )\over{\partial \rho}}{d\rho\over{d\rho_{\rm B}}}\right)
\nonumber\\
&=&{k_{\rm F}^2\over{3\rho E_{\rm F}^*}}+{m^*\over{E_{\rm F}^*}}
\left(-g_{\rm s\sigma}^*{d\sigma\over{d\rho_{\rm B}}}
-g_{\rm s\omega}^*{d\omega\over{d\rho_{\rm B}}}
-g_{\rm s\rho}^*{d\rho\over{d\rho_{\rm B}}}\right)
\nonumber\\
&+&
\left(g_{\rm v\sigma}^*{d\sigma\over{d\rho_{\rm B}}}
+g_{\rm v\omega}^*{d\omega\over{d\rho_{\rm B}}}
+g_{\rm v\rho}^*{d\rho\over{d\rho_{\rm B}}}\right)
\nonumber\\
&=&{k_{\rm F}^2\over{3\rho_{\rm B} E_{\rm F}^*}}+^t{\hat{\bf g}}
{d{\bf \Phi}\over{d\rho_{\rm B}}}
={k_{\rm F}^2\over{3\rho_{\rm B}E_{\rm F}^*}}-^t{\hat{\bf g}} ({M^*}^2)^{-1}{\hat{\bf g}}. 
\label{eq:E65}
\end{eqnarray}
%%%%%%%%
Therefore, we get the relation among the effective masses, the effective couplings and the incompressibility. \cite{rf:Tuchitani,rf:Kouno} 
%%%%%%%%
\begin{equation}
K=9\rho_{\rm B0}\left.\left({k_{\rm F}^2\over{3\rho_{\rm B} E_{\rm F}^*}}-^t{\hat{\bf g}}({M^*}^2)^{-1}{\hat{\bf g}}\right)\right|_
{\scriptstyle\rho_{\rm B} =\rho_{\rm B0} \atop\scriptstyle\rho_3=0}. 
\label{eq:E66}
\end{equation}
%%%%%%%%
In particular, if $({M^*}^2)^{-1}$ is diagonal, 
%%%%%%%%
\begin{equation}
-^t{\hat{\bf g}}({M^*}^2)^{-1}{\hat{\bf g}}
={{\hat{g}_{\omega}}^2\over{{m_{\omega}^*}^2}}
-{{\hat{g}_{\sigma}}^2\over{{m_{\sigma}^*}^2}}{\hat{\gamma}}^{-2}.
\label{eq:E67} 
\end{equation}
%%%%%%%%
Therefore, this quantity represents the difference between the strengths of the effective repulsive force and the effective attractive force. 
For the well-known parameter sets,\cite{rf:Reinhard,rf:Sharma,rf:Sugahara,rf:Lalazissis} this quantity almost vanishes at $\rho_{\rm B}=\rho_{\rm B0}$ and $\rho_3=0$. \cite{rf:Tuchitani,rf:Kouno} 

Similarly, the symmetry energy $a_4$ is given by
%%%%%%%%
\begin{eqnarray}
a_4&\equiv&{1\over{2}}\rho_{\rm B0}\left.{d^2\epsilon\over{d\rho_3}}\right|_{\scriptstyle\rho_{\rm B} =\rho_{\rm B0} \atop\scriptstyle\rho_3=0}
={k_{\rm F0}^2\over{6E_{\rm F0}}}-{1\over{2}}\rho_{\rm B0}{d\Sigma_{\rm r}(\sigma,\omega,\rho )\over{d\rho_3}}
\nonumber\\
&=&
{k_{\rm F0}^2\over{6E_{\rm F0}}}-{1\over{2}}\rho_{\rm B0}
\left({\partial\Sigma_{\rm r}\over{\partial \sigma}}{d\sigma\over{d\rho_3}}
+{\partial\Sigma_{\rm r}\over{\partial \omega}}{d\omega\over{d\rho_3}}
+{\partial\Sigma_{\rm r}\over{\partial \rho}}{d\rho\over{d\rho_3}}\right)
\nonumber\\
&=&
{k_{\rm F0}^2\over{6E_{\rm F0}^*}}+{1\over{2}}{\rho_{\rm B0}}^t{\tilde{\bf g}}{d{\bf \Phi}\over{d\rho_3}}
\nonumber\\
&=&
{k_{\rm F0}^2\over{6E_{\rm F0}^*}}-{1\over{2}}{\rho_{\rm B0}}^t{\tilde{\bf g}}({M^*}^2)^{-1}{\tilde{\bf g}}, 
\label{eq:E68}
\end{eqnarray}
%%%%%%%%
where $k_{\rm F0}$ is the Fermi momentum at the normal density and $E_{\rm F0}^*=\sqrt{k_{\rm F0}^2+{m^*}^2}$. 
We have used Eq. (\ref{eq:E59}) in Eq. (\ref{eq:E68}).  

If there is a higher order of the $\omega$-meson interaction, the value of the $\omega$-meson mean field may not be proportional to the baryon density. 
However, in a viewpoint of the quark ($q$) physics, it is natural that the value of the $\omega$-meson mean field is proportional to the baryon density $\rho_{\rm B}$, because $\omega$ is related to the quark number density $\bar{q}\gamma_0q$. \cite{rf:Kunihiro} 
In this case, the off-diagonal parts ${m_{\sigma\omega}^*}^2$ and ${m_{\omega\rho}^*}^2$ vanish and Eq. (\ref{eq:E45}) yields 
%%%%%%%%
\begin{eqnarray}
{{m_\omega^*}^2\over{m_\omega^2}}={\hat{g}_\omega\over{g_\omega}}
={g_{\rm v\omega}^*-\hat{\gamma}^{-1}g_{\rm s\omega}^*\over{g_\omega}}. 
\label{eq:E68a}
\end{eqnarray}
%%%%%%%%
Therefore, the effective $\omega$-nucleon coupling automatically decreases, if the effective $\omega$-meson mass decreases. 
Combining Eqs. ({\ref{eq:E67}) with (\ref{eq:E68a}), we obtain 
%%%%%%%%
\begin{equation}
-^t{\hat{\bf g}}({M^*}^2)^{-1}{\hat{\bf g}}
={{\hat{g}_{\omega}}^2\over{{m_{\omega}^*}^2}}
-{{\hat{g}_{\sigma}}^2\over{{m_{\sigma}^*}^2}}{\hat{\gamma}}^{-2}.
={\hat{g}_{\omega}g_\omega\over{m_\omega^2}}
-{{\hat{g}_{\sigma}}^2\over{{m_{\sigma}^*}^2}}{\hat{\gamma}}^{-2}.
\label{eq:E67a} 
\end{equation}
%%%%%%%%
The right hand side of equation (\ref{eq:E67a}) means that the EOS becomes softer as $m^*_\omega$ becomes smaller, since $\hat{g}_\omega$ decreases according to (\ref{eq:E68a}). 

Similarly, if we require that the value of the $\rho$-meson mean field is proportional to the isovector density $\rho_3$, we obtain 
%%%%%%%%
\begin{eqnarray}
{{m_\rho^*}^2\over{m_\rho^2}}={\tilde{g}_\rho\over{g_\rho}}={g_{\rm r\rho}^*-\tilde{\gamma}^{-1}g_{\rm s\rho}^*\over{g_\rho}}
\label{eq:E68b}
\end{eqnarray}
%%%%%%%%
and 
%%%%%%%%
\begin{equation}
-^t{\tilde{\bf g}}({M^*}^2)^{-1}{\tilde{\bf g}}
={{\tilde{g}_{\rho}}^2\over{{m_{\rho}^*}^2}}
-{{\tilde{g}_{\sigma}}^2\over{{m_{\sigma}^*}^2}}{\tilde{\gamma}}^{-2}.
={\tilde{g}_{\rho}g_\rho\over{m_\rho^2}}
-{{\tilde{g}_{\sigma}}^2\over{{m_{\sigma}^*}^2}}{\tilde{\gamma}}^{-2}.
\label{eq:E67b} 
\end{equation}
%%%%%%%%
Therefore, the EOS becomes softer if the the effective $\rho$-meson mass decreases, since the effective $\rho$-nucleon coupling $\tilde{g}_\rho$ decreases according to (\ref{eq:E68b}). 

%%%%%%%%%%%%%%%%%%%%%%%%%%%%%%%%%%%%%%%%%%%%%%%%%%%%%%%%%%%%%%%%%%%%%%%%%%%%
\section{Effective hadron masses, effective coupling and equation of state}
%%%%%%%%%%%%%%%%%%%%%%%%%%%%%%%%%%%%%%%%%%%%%%%%%%%%%%%%%%%%%%%%%%%%%%%%%%%%

In this section, we numerically investigate the EOS in asymmetric nuclear matter. 
Below we assume that 
%%%%%%%
\begin{eqnarray}
\Sigma_{\rm s}(\sigma )&=&-g_\sigma\sigma , 
\label{eq:E69}\\
\Sigma_{\rm v} (\sigma, \omega )&=&-g_\omega\omega -g_{\omega 3}\omega^3
-g_{\sigma 2\omega}\sigma^2\omega, 
\label{eq:E70}\\
\Sigma_{\rm r} (\sigma, \rho )&=&-g_\rho\rho -g_{\rho 3}\rho^3-g_{\sigma 2\rho}\sigma^2\rho
,
\label{eq:E71}
\end{eqnarray}
%%%%%%%
and 
%%%%%%%
\begin{eqnarray}
U_{\rm M}(\sigma ,\omega, \rho )&=&{1\over{2}}m_{\sigma}^2\sigma^2+{1\over{3}}c_{\sigma 3}\sigma^3+{1\over{4}}c_{\sigma 4}\sigma^4
\nonumber\\
&-&{1\over{2}}m_{\omega}^2\omega^2-{1\over{4}}c_{\omega 4}\omega^4
+{1\over{2}}c_{\sigma 2\omega 2}\sigma^2\omega^2
\nonumber\\
&-&{1\over{2}}m_\rho^2\rho^2
-{1\over{4}}c_{\rho 4}\rho^4+{1\over{2}}c_{\sigma 2\rho 2}\sigma^2\rho^2.
\label{eq:E72}
\end{eqnarray}
%%%%%%%
Under these assumptions, 
the effective meson-nucleon couplings are given by 
%%%%%%%%
\begin{eqnarray}
{\bf g}_{\rm s}^*
=\left[
\begin{array}{c}
g_{\rm s\sigma}^* \\
g_{\rm s\omega}^* \\
g_{\rm s\rho  }^* 
\end{array}
\right]
=\left[
\begin{array}{c}
g_\sigma \\
0 \\
0
\end{array}
\right]
\label{eq:E73}
\end{eqnarray}
%%%%%%%%
%%%%%%%%
\begin{eqnarray}
{\bf g}_{\rm v}^*
=\left[
\begin{array}{c}
g_{\rm v\sigma}^* \\
g_{\rm v\omega}^* \\
g_{\rm v\rho  }^* 
\end{array}
\right]
=\left[
\begin{array}{c}
2g_{\sigma 2\omega}\sigma\omega \\
g_\omega+3g_{\omega 3}\omega^2+g_{\sigma 2\omega}\sigma^2 \\
0
\end{array}
\right]
\label{eq:E74}
\end{eqnarray}
%%%%%%%%
and 
%%%%%%%%
\begin{eqnarray}
{\bf g}_{\rm r}^*
=\left[
\begin{array}{c}
g_{\rm r\sigma}^* \\
g_{\rm r\omega}^* \\
g_{\rm r\rho  }^* 
\end{array}
\right]
=\left[
\begin{array}{c}
2g_{\sigma 2\rho}\sigma\rho \\
0 \\
g_\rho +3g_{\rho 3}\rho^2+g_{\sigma 2\rho}\sigma^2
\end{array}
\right].
\label{eq:E75}
\end{eqnarray}
%%%%%%%%
Since $g^*_{\rm s\omega}=g^*_{\rm s\rho}=0$, $\hat{g}_\omega =g_\omega^*$ and $\tilde{g}_\rho=g_\rho^*$ in this model. 

The effective meson masses are given by following equations. 
%%%%%%%%
\begin{eqnarray}
{m^*_\sigma}^2&=&{m_\sigma}^2+2c_{\sigma 3}\sigma+3c_{\sigma 4}\sigma^2
\nonumber\\
&+&{{g_{\rm s\sigma}^*}^2\over{2\pi^2}}\left\{k_{\rm Fp}E_{\rm Fp}^*+2{k_{\rm Fp}{m^*}^2\over{{E_{\rm Fp}^*}}}-3{m^*}^2\log{\left({k_{\rm Fp}+E_{\rm Fp}^*\over{m^*}}\right)}\right\}
\nonumber\\
&+&{{g_{\rm s\sigma}^*}^2\over{2\pi^2}}\left\{k_{\rm Fn}E_{\rm Fn}^*+2{k_{\rm Fn}{m^*}^2\over{{E_{\rm Fn}^*}}}-3{m^*}^2\log{\left({k_{\rm Fn}+E_{\rm Fn}^*\over{m^*}}\right)}\right\}
\nonumber\\
&+&2g_{\sigma 2\omega }\omega\rho_{\rm B}+c_{\sigma 2\omega 2}\omega^2
+2g_{\sigma 2\rho }\rho\rho_{3}+c_{\sigma 2\rho 2}\rho^2. 
\label{eq:E76}
\\
{m^*_\omega}^2&=&{m_\omega}^2+3c_{\omega 4}\omega^2
-6g_{\omega 3}\omega\rho_{\rm B}-c_{\sigma 2\omega 2}\sigma^2, 
\label{eq:E77}
\\
{m^*_\rho}^2&=&{m_\rho}^2+3c_{\rho 4}\rho^2
-6g_{\rho 3}\rho\rho_3-c_{\sigma 2\rho 2}\sigma^2. 
\label{eq:E78}
\end{eqnarray}
%%%%%%%%
The off-diagonal elements of the matrix ${M^*}^2$ vanish except for 
%%%%%%%%
\begin{eqnarray}
{m_{\sigma\omega}^*}^2&=&2c_{\sigma 2\omega 2}\sigma\omega +2g_{\sigma 2\omega}\sigma\rho_B
+2c_{\sigma_2\rho 2}\sigma\rho+2g_{\sigma 2\rho}\sigma\rho_3. 
\label{eq:E79}
\end{eqnarray}
%%%%%%%%

For numerical calculations, we use the parameter set (P.S.) NL3 \cite{rf:Lalazissis} as a basic one. 
In the P.S. NL3, all parameters except for $m_\sigma$, $m_\omega$, $m_\rho$, $g_\sigma$, $g_\omega$, $g_\rho$, $c_{\sigma 3}$ and $c_{\sigma 4}$ vanish. 
In infinite nuclear matter, the parameters $g_\sigma$ ($g_\omega$, $g_\rho$) and $m_\sigma$ ($m_\omega$, $m_\rho$) appear in the theory, only through the form of $g_\sigma /m_\sigma$ ($g_\omega /m_\omega$, $g_\rho /m_\rho$). 
Therefore, we have five independent parameters in this case. 
We add one (or two) fixed parameter(s) of the higher-order interaction(s) to the five basic parameters and re-determine the five parameters to reproduce the five basic properties of the EOS for the P.S. NL3, namely, $\rho_{\rm B0}=0.148$fm$^{-3}$, $a_1=16.299$MeV, $m^*_0=0.6m$, $K=271.76$MeV and $a_4=37.4$MeV. 
(We also put $m=939$MeV as in the case of the P.S. NL3.)

In a P.S. B, we add the $\omega$-meson self-interaction $\omega^4$ to the basic one. 
This term makes the effective $\omega$-meson mass larger (smaller) if $c_{\omega 4}$ is positive (negative). 
An example of this parameter set is shown in Table 1.  
In the P.S. B$_1$ , $c_{\omega 4}$ is chosen to yield ${m_\omega^*}^2/m_\omega^2=$0.9 at $\rho_{\rm B}=\rho_{B0}$ and $\rho_3=0$. 

%%%%%%%%%%%%%%%%%%%%%%%% table %%%%%%%%%%%%%%%%%%%%%%%%%%%%%%%%%%%%%%%%%%%%
\begin{table}[ht]
\begin{center}
\begin{tabular}{lccccc} \hline \hline
                               & B$_1$   & D$_1$   & BD$_1$  &              \\
\hline
${g_{\sigma}}^2/{m_\sigma}^2$  & 395.474 & 384.641 & 407.537 & (GeV$^{-2}$) \\  \hline 
${g_{\omega}}^2/{m_\omega}^2$  & 295.581 & 250.491 & 278.194 & (GeV$^{-2}$) \\  \hline 
${g_{\rho}}^2/{m_\rho}^2$      & 34.9375 & 34.9375 & 34.9375 & (GeV$^{-2}$) \\  \hline 
${g_{\omega 3}}/{g_\omega}^3$  & 0.0     & 0.20    & -0.35   & (GeV$^{-2}$) \\
\hline 
${c_{\sigma 3}}/{g_\sigma}^3$  & 2.26435 & 2.38373 & 1.49380  & (MeV)        \\
\hline 
${c_{\sigma 4}}/{g_\sigma}^4$  &-3.79059 &-4.61320 &-1.20853  &$\times 10^{-3}$  \\ \hline 
${c_{\omega 4}}/{g_\omega}^4$  &-0.0014   & 0.0    &-0.0037744&              \\
\hline 
\hline
\end{tabular}
\end{center}
\caption{The P.S. B$_1$, D$_1$ and BD$_1$. All of the other parameters are zero. }
\label{Table 1}
\end{table}
%%%%%%%%%%%%%%%%%%%%%%%%%%%%%%%%%%%%%%%%%%%%%%%%%%%%%%%%%%%%%%%%%%%%%%%%%%%
 
In a P.S. D, we add the $\omega$-$\omega$-$\omega$-nucleon interaction. 
This term makes the effective $\omega$-nucleon coupling larger (smaller) and the effective $\omega$-meson mass smaller (larger) if $g_{\omega 3}$ is positive (negative). 
An example of this parameter set is shown in Table 1. 
In the P.S. D$_1$ , $g_{\omega 3}$ is chosen to yield ${m_\omega^*}^2/m_\omega^2=$0.9 at $\rho_{\rm B}=\rho_{B0}$ and $\rho_3=0$.  

If we require that the $\omega$-meson mean field is proportional to the baryon density, the relation (\ref{eq:E68a}) is satisfied. 
In a P.S. BD, we add the $\omega$-meson self-interaction $\omega^4$ and the $\omega$-$\omega$-$\omega$-nucleon interaction at the same time, and put $c_{\omega 4}=3g_{\omega 3}m_\omega^2/g_\omega$ to satisfy the relation (\ref{eq:E68a}). 
An example of this parameter set is shown in Table 1. 
In the P.S. BD$_1$, $g_{\omega 3}$ and $c_{\omega 4}$ are chosen to yield ${m_\omega^*}^2/m_\omega^2=\hat{g}_\omega/g_\omega =$0.90 at $\rho_{\rm B}=\rho_{B0}$ and $\rho_3=0$. 

In Figs. 2$\sim$6, we show the results for the P.S. B$_1$, D$_1$ and BD$_1$. 
In the P.S. B$_1$, the effective $\omega$-meson mass decreases as density increases. 
This makes $K(\rho_{\rm B})$ larger and the EOS becomes stiffer at high density. 
As is seen in Fig. 6, the EOS also becomes stiffer in the asymmetric nuclear matter.
We emphasize that this behavior is not trivial since the basic four parameters, $g_\sigma/m_\sigma$, $g_\omega/m_\omega$, $c_{\sigma 3}$ and $c_{\sigma 4}$ are also changed to reproduce the basic properties of EOS for the P.S. NL3 at the normal density, namely $\rho_{\rm B0}=0.148$fm$^{-3}$, $a_1=16.299$MeV, $m^*_0=0.6m$ and $K=271.76$MeV. 
(Since the $\omega^4$ interaction does not affect $a_4$ and $g_\rho/m_\rho$ is not changed in this case. )

In the P.S. D$_1$, the effective $\omega$-meson mass decreases and the effective $\omega$-nucleon coupling increases as density increases. 
This makes $K(\rho_{\rm B})$ much larger and the EOS becomes much stiffer at high density. 
As is seen in Fig. 6, the EOS also becomes much stiffer in the asymmetric nuclear matter.

In the P.S. BD$_1$, the effective $\omega$-meson mass decreases and the effective $\omega$-nucleon coupling decreases as density increases. 
As is seen in the previous section, 
the effect of the reduction of the effective $\omega$-nucleon coupling overcomes the effect of the reduction of the effective $\omega$-meson mass. 
Therefore, $K(\rho_{\rm B})$ becomes smaller and the EOS becomes softer at high density. 
As is seen in Fig. 6, the EOS also becomes softer in the asymmetric nuclear matter.

In a P.S. E, we add the $\rho$-meson self-interaction $\rho^4$. 
This term makes the effective $\rho$-meson mass larger (smaller) at large $|\rho_3|$ if $c_{\rho 4}$ is positive (negative). 
An example of this parameter set is shown in Table 2. 
In the P.S. E$_1$, $c_{\rho 4}$ is chosen to yield ${m_\rho^*}^2/m_\rho^2=$0.9 at $\rho_{\rm B}=\rho_{\rm B0}$ and $\rho_3=-\rho_{\rm B0}$. 

%%%%%%%%%%%%%%%%%%%%%%%% table %%%%%%%%%%%%%%%%%%%%%%%%%%%%%%%%%%%%%%%%%%%%
\begin{table}[ht]
\begin{center}
\begin{tabular}{lccccc} \hline \hline
                               & E$_1$   & F$_1$   & EF$_1$  &              \\
\hline
${g_{\sigma}}^2/{m_\sigma}^2$  & 402.857 & 402.857 & 402.857 & (GeV$^{-2}$) \\  \hline 
${g_{\omega}}^2/{m_\omega}^2$  & 268.597 & 268.597 & 268.597 & (GeV$^{-2}$) \\  \hline 
${g_{\rho}}^2/{m_\rho}^2$      & 34.9375 & 34.9375 & 34.9375 & (GeV$^{-2}$) \\  \hline 
${g_{\rho 3}}/{g_\rho}^3$      & 0.0     & 10.1313 &-21.329  & (GeV$^{-2}$) \\
\hline 
${c_{\sigma 3}}/{g_\sigma}^3$  & 1.95823 & 1.95823 & 1.95823 & (MeV)        \\
\hline 
${c_{\sigma 4}}/{g_\sigma}^4$  &-2.64707 &-2.64707 &-2.64707 &$\times 10^{-3}$  \\ \hline 
${c_{\rho 4}}/{g_\rho}^4$      &-0.57047 & 0.0     &-1.83147 &              \\
\hline 
\hline
\end{tabular}
\end{center}
\caption{The P.S. E$_1$, F$_1$ and EF$_1$. All of the other parameters are zero. }
\label{Table 2}
\end{table}
%%%%%%%%%%%%%%%%%%%%%%%%%%%%%%%%%%%%%%%%%%%%%%%%%%%%%%%%%%%%%%%%%%%%%%%%%%%

In a P.S. F, we add the $\rho$-$\rho$-$\rho$-nucleon interaction. 
This term makes the effective $\rho$-nucleon coupling larger (smaller) and the effective $\rho$-meson mass smaller (larger) at large $|\rho_3|$ if $g_{\rho 3}$ is positive (negative). 
An example of this parameter set is shown in Table 2. 
In the P.S. F$_1$, $g_{\rho 3}$ is chosen to yield ${m_\rho^*}^2/m_\rho^2=$0.9 at $\rho_{\rm B}=\rho_{\rm B0}$ and $\rho_3=-\rho_{\rm B0}$. 

If we require that the $\rho$-meson mean field is proportional to the isovector density, the relation (\ref{eq:E68b}) is satisfied. 
In a P.S. EF, we add the $\rho$-meson self-interaction $\rho^4$ and the $\rho$-$\rho$-$\rho$-nucleon interaction at the same time, and put $c_{\rho 4}=3g_{\rho 3}m_\rho^2/g_\rho$ to satisfy the relation (\ref{eq:E68b}). 
An example of this parameter set is shown in Table 2. 
In the P.S. EF$_1$, $g_{\rho 3}$ and $c_{\sigma 2\rho 2}$ are chosen to yield ${m_\rho^*}^2/m_\rho^2=\tilde{g}_\rho/g_\rho =$0.90 at $\rho_{\rm B}=\rho_{B0}$ and $\rho_3=-\rho_{\rm B0}$. 
 
In Figs. 7$\sim$9, we show the results for the P.S. E$_1$, F$_1$ and EF$_1$. 
In the P.S. E$_1$, the effective $\rho$-meson mass decreases as $|\rho_3|$ increases. 
This makes the EOS becomes stiffer at large $|\rho_3|$. 

In the P.S. F$_1$, the effective $\rho$-meson mass decreases and the effective $\rho$-nucleon coupling increases at large $|\rho_3|$. 
This makes the EOS becomes much stiffer at large $|\rho_3|$. 

In the P.S. EF$_1$, the effective $\rho$-meson mass decreases and the effective $\rho$-nucleon coupling decreases as $|\rho_3|$ increases. 
As is seen the previous section, 
the effect of the reduction of the effective $\rho$-nucleon coupling overcomes the effect of the reduction of the effective $\rho$-meson mass. 
Therefore, the EOS becomes softer at large $|\rho_3|$. 

Finally, we consider the meson-mixing interactions. 
In a P.S. KL, we add the mixing interaction $\sigma^2\omega^2$ and the $\sigma$-$\sigma$-$\omega$-nucleon interaction at the same time, and put $c_{\sigma 2\omega 2}=-g_{\sigma 2\omega}m_\omega^2/g_\omega$ to satisfy the relation (\ref{eq:E68a}). 
Although there is the $\sigma$-$\omega$ mixing interactions, the off-diagonal element ${m_{\sigma\omega}^*}^2$ vanishes because of this condition. 
In this model, the effective $\omega$-meson mass larger (smaller) and the effective $\omega$-nucleon coupling larger (smaller) if $g_{\sigma 2\omega}$ is positive (negative). 
The effective $\omega$-nucleon coupling and the effective $\omega$-meson mass are given by 
%%%%%%%
\begin{eqnarray}
{\hat{g}_{\omega}\over{g_\omega}}={{m_\omega^*}^2\over{m_\omega^2}}=1+{g_{\sigma 2\omega}\over{g_\omega}}\sigma^2=1+{g_{\sigma 2\omega}\over{g_\sigma^2g_\omega}}(m-m^*)^2. 
\label{eq:E100}
\end{eqnarray}
%%%%%%%
An example of this parameter set is shown in Table 3. 
In the P.S. KL$_1$, $g_{\sigma 2\omega}$ and $c_{\sigma 2\omega 2}$ are chosen to yield ${m_\omega^*}^2/m_\omega^2=\hat{g}_\omega/g_\omega =$0.90 at $\rho_{\rm B}=\rho_{\rm B0}$ and $\rho_3=0$. 

%%%%%%%%%%%%%%%%%%%%%%%% table 3 %%%%%%%%%%%%%%%%%%%%%%%%%%%%%%%%%%%%%%%%%%%%
\begin{table}[ht]
\begin{center}
\begin{tabular}{lccccc} \hline \hline
                               & KL$_1$  & GH$_1$  & KLGH$_1$&              \\
\hline
${g_{\sigma}}^2/{m_\sigma}^2$  & 408.915 & 402.857 & 408.915 & (GeV$^{-2}$) \\  \hline 
${g_{\omega}}^2/{m_\omega}^2$  & 298.440 & 268.597 & 298.440 & (GeV$^{-2}$) \\  \hline 
${g_{\rho}}^2/{m_\rho}^2$      & 34.9375 & 38.8194 & 38.8194 & (GeV$^{-2}$) \\  \hline 
${g_{\sigma 2\omega}}/({g_\sigma}^2g_\omega)$&-0.70883 & 0.0     &-0.70883  & (GeV$^{-2}$) \\
\hline 
${g_{\sigma 2\rho}}/({g_\sigma}^2g_\rho)    $& 0.0     &-0.70883 &-0.70883  & (GeV$^{-2}$) \\
\hline 
${c_{\sigma 3}}/{g_\sigma}^3$  & 0.0886639  & 1.95823  & 0.0886639 & (MeV)        \\
\hline 
${c_{\sigma 4}}/{g_\sigma}^4$  & 4.51100 &-2.64707 & 4.51100 &$\times 10^{-3}$  \\ \hline 
${c_{\sigma 2\omega 2}}/({g_\sigma}^2{g_\omega}^4)$  & 2.37511 & 0.0     &  2.37511       & $\times 10^{-3}$   \\
\hline 
${c_{\sigma 2\rho 2}}/({g_\sigma}^2{g_\rho}^2)$ & 0.0  & 1.82597 & 1.82597 &    $\times 10^{-2}$          \\
\hline 
\hline
\end{tabular}
\end{center}
\caption{The P.S. KL$_1$,GHF$_1$ and KLGH$_1$. All of the other parameters are zero.}
\label{Table 3}
\end{table}
%%%%%%%%%%%%%%%%%%%%%%%%%%%%%%%%%%%%%%%%%%%%%%%%%%%%%%%%%%%%%%%%%%%%%%%%%%%
 
In a P.S. GH, we add the mixing interaction $\sigma^2\rho^2$ and the $\sigma$-$\sigma$-$\rho$-nucleon interaction at the same time, and put $c_{\sigma 2\rho 2}=-g_{\sigma 2\rho}m_\rho^2/g_\rho$ to satisfy the relation (\ref{eq:E68b}). 
Although there is the $\sigma$-$\rho$ mixing interactions, the off-diagonal element ${m_{\sigma\rho}^*}^2$ vanishes because of this condition. 
In this model, the effective $\rho$-meson mass larger (smaller) and the effective $\rho$-nucleon coupling larger (smaller) at large $|\rho_3|$ if $g_{\sigma 2\rho}$ is positive (negative). 
The effective $\rho$-nucleon coupling and the effective $\rho$-meson mass are given by 
%%%%%%%
\begin{eqnarray}
{\tilde{g}_{\rho}\over{g_\rho}}={{m_\rho^*}^2\over{m_\rho^2}}=1+{g_{\sigma 2\rho}\over{g_\rho}}\sigma^2=1+{g_{\sigma 2\rho}\over{g_\sigma^2g_\rho}}(m-m^*)^2.
\label{eq:E101}
\end{eqnarray}
%%%%%%%
An example of this parameter set is shown in Table 3. 
In the P.S. GH$_1$, $g_{\sigma 2\rho}$ and $c_{\sigma 2\rho 2}$ are chosen to yield ${m_\rho^*}^2/m_\rho^2=\tilde{g}_\rho/g_\rho =$0.90 at $\rho_{\rm B}=\rho_{\rm B0}$ and $\rho_3=0$. 

In a P.S. KLGH, we add the mixing interactions $\sigma^2\omega^2$, the $\sigma$-$\sigma$-$\omega$-nucleon interaction, the mixing interaction $\sigma^2\rho^2$ and the $\sigma$-$\sigma$-$\rho$-nucleon interaction at the same time, and put $c_{\sigma 2\omega 2}=-g_{\sigma 2\omega}m_\omega^2/g_\omega$ and $c_{\sigma 2\rho 2}=-g_{\sigma 2\rho}m_\rho^2/g_\rho$ to satisfy the relations (\ref{eq:E68a}) and  (\ref{eq:E68b}). 
We also put $g_{\sigma 2\omega}/g_\omega =g_{\sigma 2\rho}/g_\rho$ to satisfy $m_\omega^*/m_\omega=m_\rho^*/m_\rho$. 
These terms makes the effective $\omega$ and $\rho$-meson masses larger (smaller) and the effective $\omega$ and $\rho$-nucleon couplings larger (smaller) if $g_{\sigma 2\omega}$ and $g_{\sigma 2\rho}$ are positive (negative). 
An example of this parameter set is shown in Table 3. 
In the P.S. KLGH$_1$, $g_{\sigma 2\omega}$, $g_{\sigma 2\rho}$, $c_{\sigma 2\omega 2}$ and $c_{\sigma 2\rho 2}$ are chosen to yield ${m_\omega^*}^2/m_\omega^2 =\hat{g}_\omega/g_\omega={m_\rho^*}^2/m_\rho^2=\tilde{g}_\rho/g_\rho =$0.90 at $\rho_{\rm B}=\rho_{\rm B0}$ and $\rho_3=0$. 

In Figs. 10$\sim$14, we show the results for the P.S. KL$_1$, GH$_1$ and KLGH$_1$. 
In the P.S. KL$_1$, the effective $\omega$-nucleon coupling decreases as the effective $\omega$-meson mass decreases. 
At any density, $\hat{g}_\omega /g_\omega={m_\omega^*}^2/m_\omega^2$ is satisfied. 
This makes $K(\rho_{\rm B})$ smaller and the EOS softer. 

In the P.S. GH$_1$, the effective $\rho$-nucleon coupling decreases as the effective $\rho$-meson mass decreases. 
At any density, $\tilde{g}_\rho /g_\rho={m_\rho^*}^2/m_\rho^2$ is satisfied. 
However, this effect makes no change in the EOS of symmetric nuclear matter, since the mean field of $\rho$-meson is zero in the symmetric nuclear matter. 
On the other hand, this effect makes the EOS softer at large $|\rho_3|$ as is seen in Fig. 13, although $m_\rho^*$ depends on $\rho_3$ only slightly. 
(See Fig. 12.) 

In the P.S. KLGH$_1$, the EOS becomes softer at large $\rho_{\rm B}$ and/or large $|\rho_3|$, since the effective $\omega$-nucleon and $\rho$-nucleon couplings decreases. 
The decrease of the effective $\rho$-meson mass is smaller at high density than in the case with the P.S. GH$_1$, because the value of the $\sigma$-meson mean field is smaller than the one in the case with the P.S. GH$_1$. 
As is seen before, in the models with the P.S. KL, GH and KLGH, the decrease of the square of the vector meson mass is proportional to $(g_\sigma\sigma)^2=(m-m^*)^2$. 

In Fig. 14, we show the relation between the effective nucleon mass and the effective vector meson masses, using the P.S. KL$_1$, GH$_1$ and KLGH$_1$. 
All of three models have the same relation between ${m_\omega^*}/m_\omega$ (and/or $m_\rho^*/m_\rho$) and $m^*/m$, because the coefficient $g_{\sigma 2\omega}/(g_\sigma^2g_\omega )$ (and/or $g_{\sigma 2\rho}/(g_\sigma^2g_\rho )$) has the same value in (\ref{eq:E100}) (and/or (\ref{eq:E101})). 
As density increases, both of the effective nucleon mass and the effective vector ($\omega$ and/or $\rho$) meson mass decreases. 
The relation is somewhat different from the Brown-Rho scaling where the effective vector meson mass is approximately proportional to the effective nucleon mass. \cite{rf:Brown} 

Modification of the EOS for the asymmetric nuclear matter may affects the properties of the neutron star. 
Using these EOS and the Tolman-Oppenheimer-Volkoff (TOV) equation, we calculate the radius of the neutron star whose mass is 1.35 times of the solar mass. 
All EOS we have used satisfy the basic properties of the EOS for the P.S. NL3 at the normal density, namely $\rho_{\rm B0}=0.148$fm$^{-3}$, $a_1=16.299$MeV, $m^*_0=0.6m$, $K=271.76$MeV and $a_4=37.4$MeV. 
We change the value(s) of the parameter(s) of the higher-order interaction(s) and re-determine the five basic parameters to satisfy the five basic properties of the EOS for the P.S. NL3. 
For the constituents of the matter in the neutron star, we only consider protons, neutrons and electrons. 
We also neglect the effects of the crust which may be important in the calculations for the lighter star. 

In Figs. 15, 16 and 17, we show the relation between the radius $R$ of the neutron star and the effective $\omega$($\rho$)-meson mass at the center of the neutron star. 
In Figs. 18, 19 and 20, we also show the relation between the radius $R$ and the effective $\omega$($\rho$)-meson mass calculated at $\rho_{\rm B}=\rho_{\rm B0}$ and $\rho_3=0$ (or $\rho_3=-\rho_{\rm B0}$) using the same parameter set. 

In Figs. 21 and 22, we show the relation between the radius $R$ and the effective $\omega$($\rho$)-nucleon coupling at the center of the neutron star. 
In Figs. 23 and 24, we also show the relation between the radius $R$ of the neutron star and the effective $\omega$($\rho$)-nucleon coupling calculated at 
$\rho_{\rm B}=\rho_{\rm B0}$ and $\rho_3=0$ (or $\rho_3=-\rho_{\rm B0}$ ) using the same parameter set. 

In the model with the P.S. B (E), the radius $R$ becomes smaller and the effective $\omega$($\rho$)-meson mass becomes larger as the coefficient $c_{\omega 4}$($c_{\rho 4}$) become larger. 
This is because the enhancement of the effective $\omega$($\rho$)-meson mass makes the EOS softer. 

In the model with the P.S. D (F), the radius $R$ becomes larger, the effective $\omega$($\rho$)-meson mass becomes smaller and the effective $\omega$($\rho$)-nucleon coupling becomes larger as the coefficient $g_{\omega 3}$ becomes larger. 
This is because both of the reduction of the effective $\omega$($\rho$)-meson mass and the enhancement of the effective $\omega$($\rho$)-nucleon coupling make the EOS stiffer. 

In the models with the P.S. BD, EF, KL, GH and KLGH, as the radius $R$ becomes smaller, the effective $\omega$(and/or $\rho$)-meson mass becomes smaller and the effective $\omega$(and/or $\rho$)-nucleon coupling becomes smaller. 
This is because the effect of the reduction of the effective $\omega$(and/or $\rho$)-nucleon coupling overcomes the effect of the reduction of the effective $\omega$(and/or $\rho$)-meson mass and makes the EOS softer. 

In the former four cases (B,D,E and F), $R$ increases as the effective vector meson mass decreases. 
On the contrary, in the latter five cases (BD,EF,KL,GH and KLGH), $R$ decreases as the effective vector meson mass decreases. 

We also see that the radius $R$ depends on the values of the effective $\omega$-meson mass and the effective $\omega$-nucleon coupling much stronger than on the values of the effective $\rho$-meson mass and the effective $\rho$-nucleon coupling. 
The ambiguity of $R$ is a few hundred meters if $\vert {m_\omega^*}^2-m_\omega^2\vert\sim 0.1m_\omega^2$ at the normal density. 

\section{Summary}

In summary, we have studied the relation between the effective meson masses, the effective meson-nucleon couplings and the EOS in the asymmetric nuclear matter. 
We have also studied the relation among the effective meson masses, the effective meson-nucleon couplings and a radius of a neutron star. 
The results obtained in this paper is summarized as follows. 

(1) The enhancement (reduction) of the effective $\omega$($\rho$)-meson mass makes the EOS softer (stiffer) at high densities (and/or at large $|\rho_3|$) even though the basic properties of the
 EOS are fixed at the normal density. 

(2) The enhancement (reduction) of the effective $\omega$($\rho$)-nucleon coupling makes the EOS stiffer (softer) at high density (and/or at large $|\rho_3|$). 

(3) If the multi $\omega$($\rho$)-nucleon interaction is added with the positive (negative) sign, this makes the effective $\omega$($\rho$)-nucleon coupling larger (smaller) and the effective $\omega$($\rho$)-meson mass smaller (larger). 
Consequently, the EOS becomes much stiffer (softer) at high densities (and/or at large $|\rho_3|$). 

(4) If we require the effective $\omega$(and/or $\rho$)-meson mass is proportional to the baryon density $\rho_{\rm B}$ (and/or the isovector density $\rho_3$), the effective $\omega$(and/or $\rho$)-nucleon coupling decreases as the effective $\omega$(and/or $\rho$)-meson mass decreases. 
The effect of the reduction of the effective vector meson-nucleon coupling overcomes the effect of the reduction of the corresponding effective vector meson mass  and the EOS becomes softer. 

(5) The value of the radius $R$ of the neutron star depends on the values of the effective $\omega$-meson mass and the effective $\omega$-nucleon coupling much stronger than on the values of the effective $\rho$-meson mass and the effective $\rho$-nucleon coupling. 
The ambiguity of $R$ is order of a few hundred meters if $\vert {m_{\omega 0}^*}^2-m_\omega^2\vert\sim 0.1m_\omega^2$ at the normal density. 

The ambiguity of order of a few hundred meters may be somewhat smaller than the change of $R$ which is predicted as a signal of the quark star. 
\cite{rf:Alford} 
In the framework of the relativistic mean field theory with exotic matters, it seems to be difficult to yield a small radius ($<10$km) for the neutron star with canonical mass. 

On the other hand, if the hadron EOS becomes softer by the modifications of the effective vector meson mass and/or the effective vector meson-nucleon coupling, the hadron phase may survive at higher density. 
It is interesting to compare the modified normal hadron EOS with the EOS of the exotic matter. 
Such investigations are now in progress. 

\bigskip

\centerline{\bf Acknowledgement}

Authors thank Prof. T. Kohmura and Dr. K. Sakamoto for useful discussions. 
One (H.K.) of the authors also thanks Prof. T. Kunihiro, Prof. K. Hagino, Prof. M. Yahiro, Prof. Y.R. Shimizu, Prof. T. Maruyama and Prof. M. Maruyama for useful discussions and suggestions. 

\vfill\eject

%%%%%%%%%%%%%%%%%%%%%%%%%%%%%%%%%%%%%%%%%%%%%%%%%%%%%%%%%%%%%%%%%%%%%%%%%%%

%%%%%%%%%%%%%%%%%%%%%%%%%%%%%%%%%%%%%%%%%%%%%%%%%%%%%%%%%%%%%%%%%%%%%%%%%%%%%%%%\vfill\eject
%%%%%%%%%%%%%%%%%%%%%%%%%%%%%%%%%%%%%%%%%%%%%%%%%%%%%%%%%%%%%%%%%%%%%%%%%%%%%%%%
%\end{document}

\bigskip

\vfill\eject

\centerline{\bf Figure Captions}

\begin{flushleft}

\bigskip

Fig. 1 Diagrammatic descriptions. 
(a) Nucleon self-energies. 
(b) Effective meson-nucleon couplings. 
(c) Equations of motion for mesons. 
(d) Meson self-energies. (Square of effective meson masses.)

\bigskip

Fig. 2 The value of ${m_\omega^*}^2/m_\omega^2$ at $\rho_3=0$ is shown as a function of the baryon density $\rho_{\rm B}$. 
The solid, dashed, dotted and dot-dashed curves represent the results for the P.S. NL3, B$_1$, D$_1$ and BD$_1$, respectively. 

\bigskip

Fig. 3 The value of ${\hat{g}_\omega}/g_\omega$ at $\rho_3=0$ is shown as a function of the baryon density $\rho_{\rm B}$. 
The solid, dashed, dotted and dot-dashed curves represent the results for the P.S. NL3, B$_1$, D$_1$ and BD$_1$, respectively. 
The dashed line for the P.S. B$_1$ coincides with the solid line for the P.S. NL3 and is constant (unity). 

\bigskip

Fig. 4 The value of $K(\rho_{\rm B})$ (in MeV) is shown as a function of the baryon density $\rho_{\rm B}$. 
The solid, dashed, dotted and dot-dashed curves represent the results for the P.S. NL3, B$_1$, D$_1$ and BD$_1$, respectively. 

\bigskip

Fig. 5 The value of $\epsilon/\rho_{\rm B}-m$ at $\rho_3=0$ (in MeV) is shown as a function of the baryon density $\rho_{\rm B}$. 
The solid, dashed, dotted and dot-dashed curves represent the results for the P.S. NL3, B$_1$, D$_1$ and BD$_1$, respectively. 

\bigskip

Fig. 6 The value of the pressure $P$ (in GeV$^{-4}$) at $\rho_{\rm B}=2\rho_{\rm B0}$ is shown as a function of the isovector density $\rho_{\rm 3}$. 
The solid, dashed, dotted and dot-dashed curves represent the results for the P.S. NL3, B$_1$, D$_1$ and BD$_1$, respectively. 

\bigskip

Fig. 7 The value of ${m_\rho^*}^2/m_\rho^2$ at $\rho_{\rm B}=2\rho_{\rm B0}$ is shown as a function of the isovector density $\rho_{\rm 3}$. 
The solid, dashed, dotted and dot-dashed curves represent the results for the P.S. NL3, E$_1$,F$_1$ and EF$_1$, respectively. 

\bigskip

Fig. 8 The value of $\tilde{g}_\rho/g_\rho$ at $\rho_{\rm B}=2\rho_{\rm B0}$ is shown as a function of the isovector density $\rho_{\rm 3}$. 
The solid, dashed, dotted and dot-dashed curves represent the results for the P.S. NL3, E$_1$,F$_1$ and EF$_1$, respectively. 
The dashed line for the P.S. E$_1$ coincides with the solid line for the P.S. NL3 and is constant (unity). 

\bigskip

Fig. 9 The value of the pressure $P$ (in GeV$^{-4}$) at $\rho_{\rm B}=2\rho_{\rm B0}$ is shown as a function of the isovector density $\rho_{\rm 3}$. 
The solid, dashed, dotted and dot-dashed curves represent the results for the P.S. NL3, E$_1$,F$_1$ and EF$_1$, respectively. 

\bigskip

Fig. 10 The values of ${m_\omega^*}^2/m_\omega^2$($=\hat{g}_\omega/g_\omega$) 
and ${m_\rho^*}^2/m_\rho^2$($=\tilde{g}_\rho/g_\rho$) at $\rho_3=0$ are shown as functions of the baryon density $\rho_{\rm B}$. 
The solid curve represents the results of ${m_\omega^*}^2/m_\omega^2$ for the P.S. NL3 and GH$_1$, and the results of ${m_\rho^*}^2/m_\rho^2$ for the P.S. NL3 and KL$_1$. 
The dotted curve represents the result of ${m_\rho^*}^2/m_\rho^2$ for the P.S. GH$_1$. 
The dot-dashed curve represents the results of ${m_\omega^*}^2/m_\omega^2$ for the P.S. KL$_1$ and KLGH$_1$, and the result of ${m_\rho^*}^2/m_\rho^2$ for the P.S. KLGH$_1$. 

\bigskip

Fig. 11 The value of $\epsilon /\rho_{\rm B}-m$ (in MeV) at $\rho_3=0$ is shown as a function of the baryon density $\rho_{\rm B}$. 
The solid curve represents the results for the P.S. NL3 and P.S. GH$_1$. 
The dot-dashed curve represents the results for the P.S. KL$_1$ and KLGH$_1$. 

\bigskip

Fig. 12 The values of ${m_\omega^*}^2/m_\omega^2$($=\hat{g}_\omega/g_\omega$) and ${m_\rho^*}^2/m_\rho^2$($=\tilde{g}_\rho /g_\rho$) at $\rho_{\rm B}=2\rho_{\rm B0}$ are shown as functions of the isovector density $\rho_3$. 
The solid curve represents the results of ${m_\omega^*}^2/m_\omega^2$ for the P.S. NL3 and GH$_1$, and the results of ${m_\rho^*}^2/m_\rho^2$ for the P.S. NL3 and  KL$_1$.  
The dotted curve represents the result of ${m_\rho^*}^2/m_\rho^2$ for the P.S. GH$_1$. 
The dot-dashed curve represents the results of ${m_\omega^*}^2/m_\omega^2$ for the P.S. KL$_1$ and KLGH$_1$, and the result of ${m_\rho^*}^2/m_\rho^2$ for the P.S. KLGH$_1$. 

\bigskip

Fig. 13 The value of the pressure $P$ (in GeV$^{-4}$) at $\rho_{\rm B}=2\rho_{\rm B0}$ is shown as a function of the isovector density $\rho_3$. 
The solid, dashed, dotted and dot-dashed curves represent the results for the P.S. NL3, KL$_1$, GH$_1$ and KLGH$_1$, respectively. 

\bigskip

Fig. 14 The value of ${m_\omega^*}/m_\omega $ for the P.S. KL$_1$ and KLGH$_1$ and the value of ${m_\rho^*}/m_\rho$ for the P.S. GH$_1$ and KLGH$_1$ are shown as functions of the effective nucleon mass $m^*/m$. 
The solid curve represents the all results.  

\bigskip

Fig. 15 The value of ${m_{\omega c}^*}^2/m_\omega^2$ at the center of the neutron star is shown as a function of the radius $R$ (in m) of the neutron star. 
The dashed, the dotted and the dot-dashed curves represent the result for the P.S. B, D and BD, respectively. 

\bigskip

Fig. 16 The value of ${m_{\rho c}^*}^2/m_\rho^2$  at the center of the neutron star is shown as a function of the radius $R$ (in m) of the neutron star. 
The dashed, the dotted and the dot-dashed curves represent the result for the P.S. E, F and EF, respectively. 

\bigskip

Fig. 17 The square of the effective vector meson masses at the center of the neutron star are shown as functions of the radius $R$ (in m) of the neutron star. 
The dashed, the dotted and the dot-dashed curves represent 
${m_{\omega c}^*}^2/m_\omega^2(={\hat{g}_{\omega c}}/g_\omega)$ for the P.S. KL, 
${m_{\rho c}^*}^2/m_\rho^2(={\tilde{g}_{\rho c}}/g_\rho)$ for the P.S. GH, 
and 
${m_{\omega c}^*}^2/m_\omega^2(={\hat{g}_{\omega c}}/g_\omega)$ and ${m_{\rho c}^*}^2/m_\rho^2(={\tilde{g}_{\rho c}}/g_\rho)$ for the P.S. KLGH, respectively. 

\bigskip

Fig. 18  The value of ${m_{\omega 0}^*}^2/m_\omega^2$ at $\rho_{\rm B}=\rho_{\rm B0}$ and $\rho_3=0$ is shown as a function of the radius $R$ (in m) of the neutron star. 
The dashed, the dotted and the dot-dashed curves represent the results for the P.S. B, D and BD, respectively. 

\bigskip

Fig. 19  The value of ${m_{\rho n}^*}^2/m_\rho^2$ calculated at $\rho_{\rm B}=\rho_{\rm B0}$ and $\rho_3=-\rho_{\rm B0}$ is shown as a function of the radius $R$ (in m) of the neutron star. 
The dashed, the dotted and the dot-dashed curves represent the result for the P.S. E, F and EF, respectively. 

\bigskip

Fig. 20  The square of the effective vector meson mass at $\rho_{\rm B}=\rho_{\rm B0}$ and $\rho_3=0$ are shown as functions of the radius $R$ (in m) of the neutron star. 
The dashed, the dotted and the dot-dashed curves represent 
the results of ${m_{\omega 0}^*}^2/m_\omega^2(={\hat{g}_{\omega 0}}/g_\omega)$ for the P.S. KL, 
${m_{\rho 0}^*}^2/m_\rho^2(={\tilde{g}_{\rho 0}}/g_\rho)$ for the P.S. GH, 
and 
${m_{\omega 0}^*}^2/m_\omega^2(={\hat{g}_{\omega 0}}/g_\omega)$ and ${m_{\rho 0}^*}^2/m_\rho^2(={\tilde{g}_{\rho 0}}/g_\rho)$ for the P.S. KLGH, respectively.  
\bigskip

Fig. 21 The value of ${\hat{g}_{\omega c}}/g_\omega$ at the center of the neutron star is shown as a function of the radius $R$ (in m) of the neutron star. 
The dashed, the dotted and the dot-dashed curves represent the result for the P.S. B, D and BD, respectively. 
The dashed line is constant (unity). 

\bigskip

Fig. 22 The value of ${\tilde{g}_{\rho c}}/g_\rho$ at the center of the neutron star is shown as a function of the radius $R$ (in m) of the neutron star. 
The dashed, the dotted and the dot-dashed curves represent the result for the P.S. E, F and EF, respectively. 
The dashed line is constant (unity). 

\bigskip

Fig. 23 The value of ${\hat{g}_{\omega 0}}/g_\omega$ at $\rho_{\rm B}=\rho_{\rm B0}$ and $\rho_3=0$ is shown as a function of the radius $R$ (in m) of the neutron star. 
The dashed, the dotted and the dot-dashed curves represent the results for the P.S. B, D and BD, respectively. 
The dashed line is constant (unity). 

\bigskip

Fig. 24  The value of ${\tilde{g}_{\rho 0n}}/g_\rho$ at $\rho_{\rm B}=\rho_{\rm B0}$ and $\rho_3=-\rho_{\rm B0}$ is shown as a function of the radius $R$ (in m) of the neutron star. 
The dashed, the dotted and the dot-dashed curves represent the result for the P.S. E, F and EF, respectively. 
The dashed line is constant (unity). 

\bigskip

\end{flushleft}

%\end{document}

%%%%%%%%%%%%%%%%%%%%%%%%%%%%%%%%%%%%%%%%%%%%%%%%%%%%%%%%%%%%%%%%%%%%%%%%%%
\vfill\eject
%%%%%%%%%%%%%%%%%%%%%%%%%%%%%%%%%%%%%%%%%%%%%%%%%%%%%%%%%%%%%%%%%%%%%%%%%%

\oddsidemargin -0mm
\textwidth 190mm
\topmargin 0mm
\textheight 250mm
\headsep -5mm
\topskip -5mm

 \begin{center}

\begin{center}
\begin{tabular}{cc}
\includegraphics[height=170mm,width=150mm] {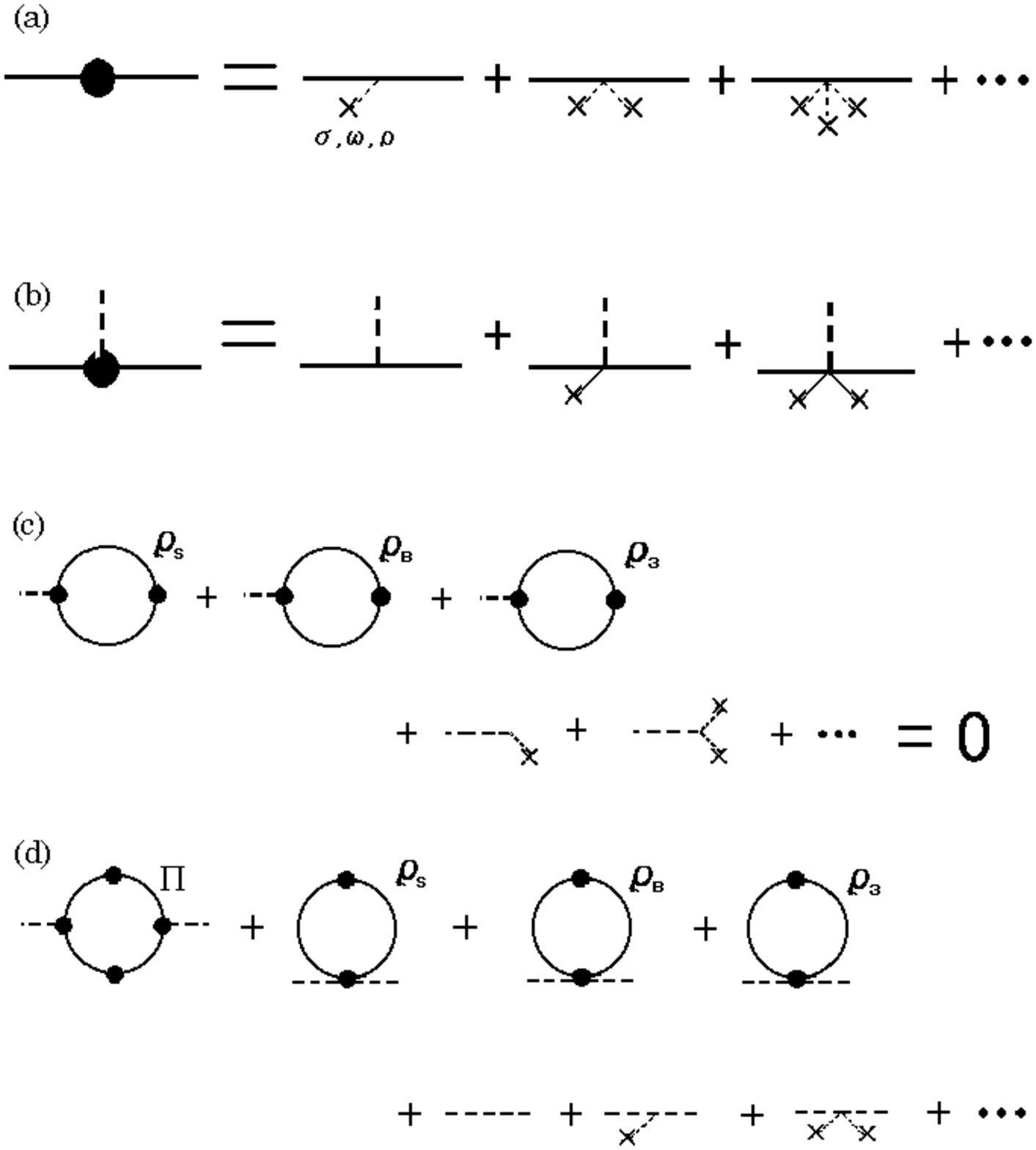} \\ 

fig.1
\end{tabular}
\end{center}

\newpage

\oddsidemargin -20mm

\begin{center}
\begin{tabular}{cc}
\includegraphics[height=70mm,width=90mm] {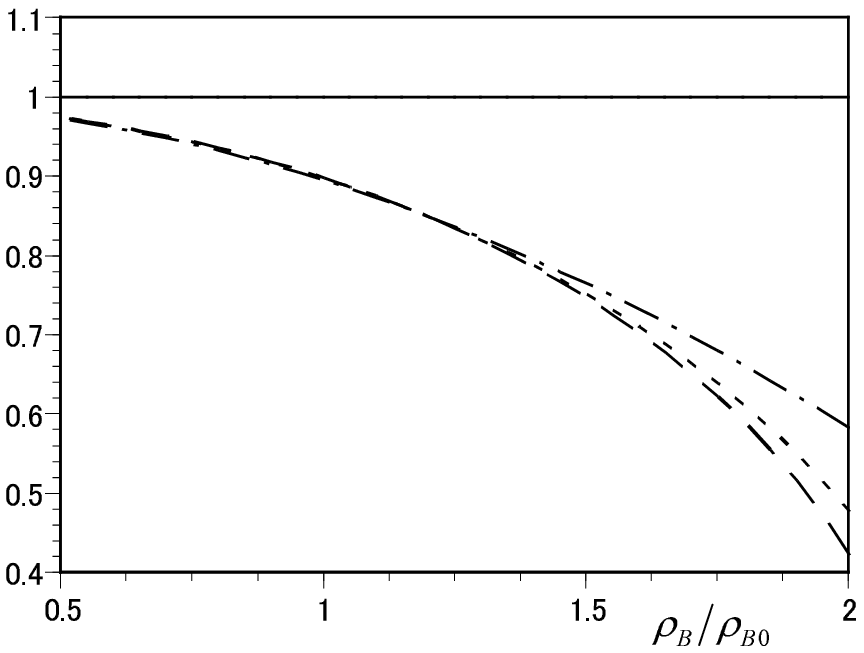} &
\includegraphics[height=70mm,width=90mm] {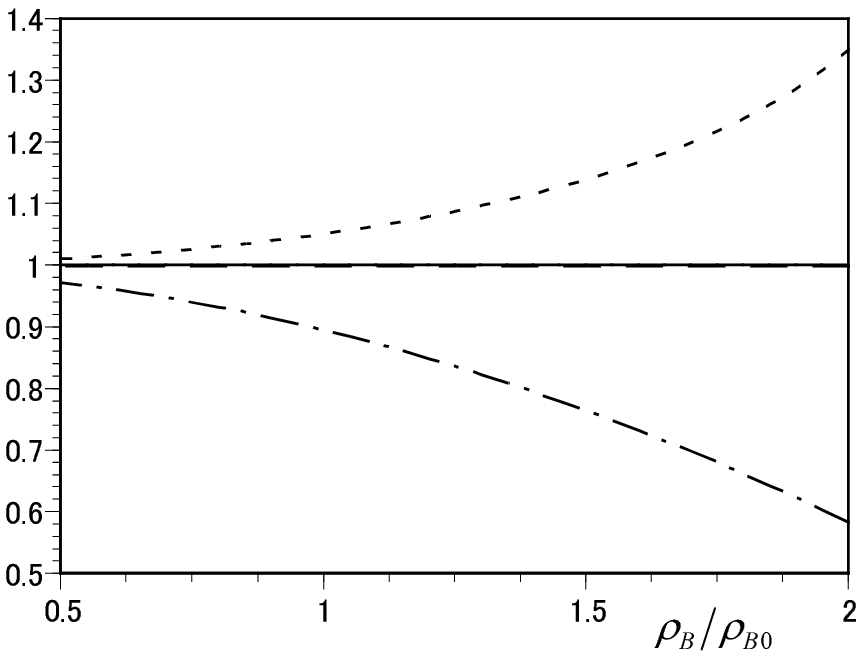}\\

fig.2 
&
fig.3
\end{tabular}
\end{center}

\begin{center}
\begin{tabular}{cc}
\includegraphics[height=70mm,width=90mm] {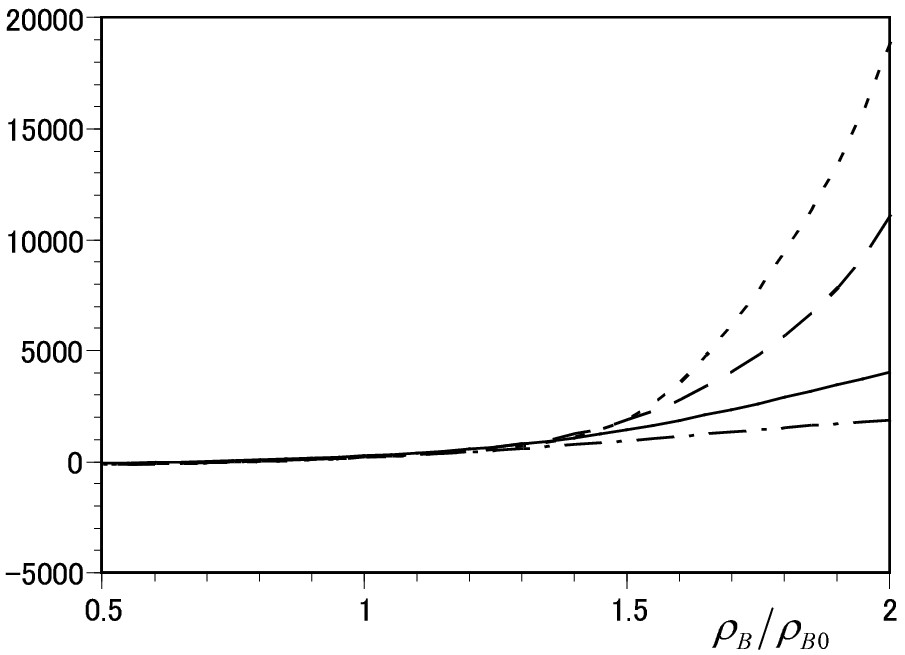}@&
\includegraphics[height=70mm,width=90mm] {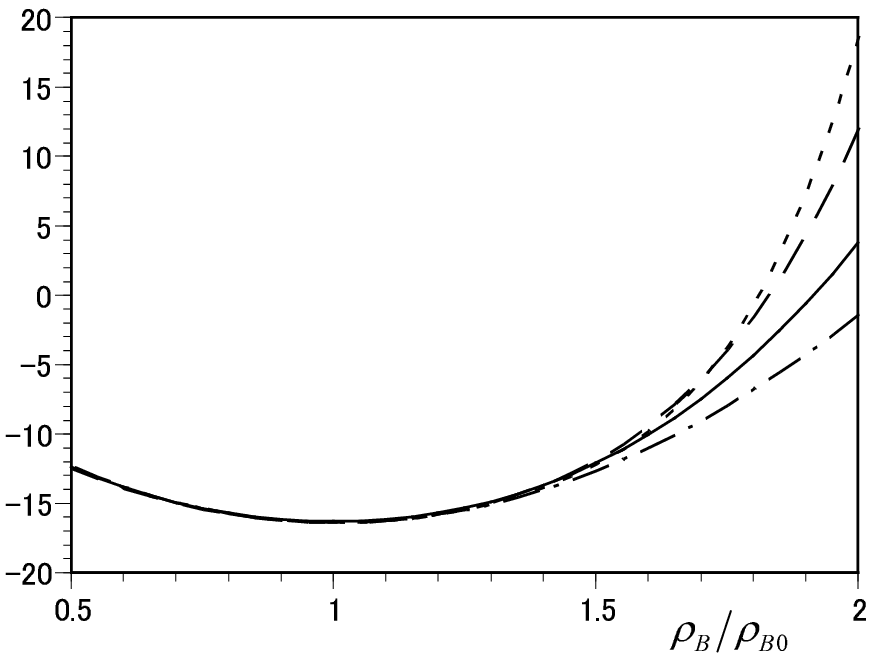} \\ 
fig.4
&
fig.5
\end{tabular}
\end{center}

\begin{center}
\begin{tabular}{cc}
\includegraphics[height=70mm,width=90mm] {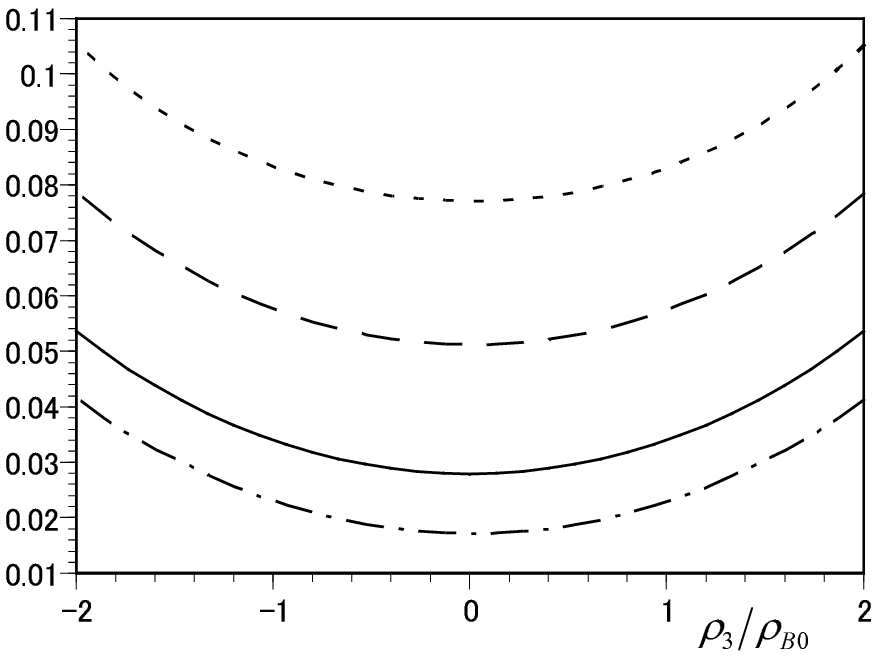}\\
fig.6
\end{tabular}
\end{center}

\begin{center}
\begin{tabular}{cc}
\includegraphics[height=70mm,width=90mm] {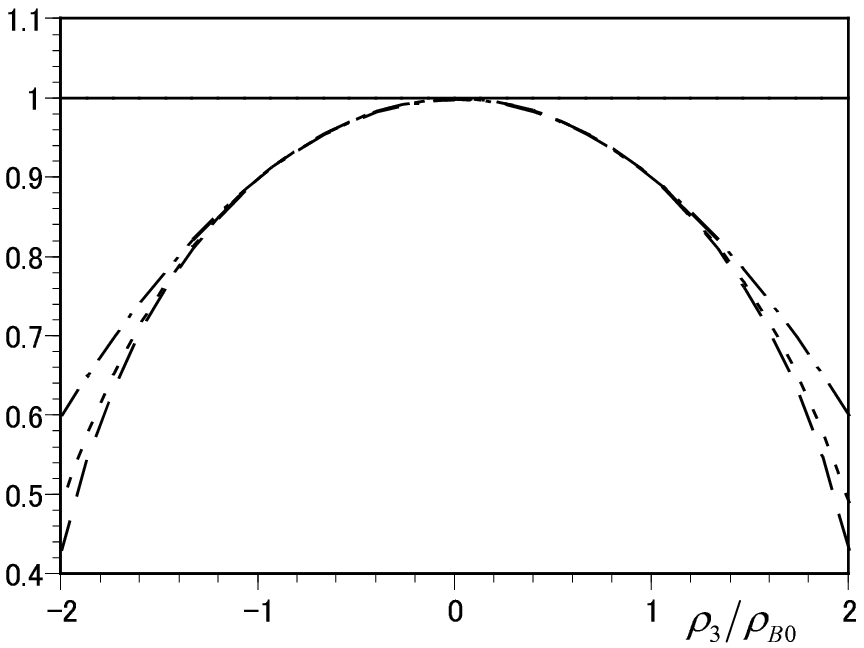}\\
fig.7
\end{tabular}
\end{center}

\begin{center}
\begin{tabular}{cc}
\includegraphics[height=70mm,width=90mm] {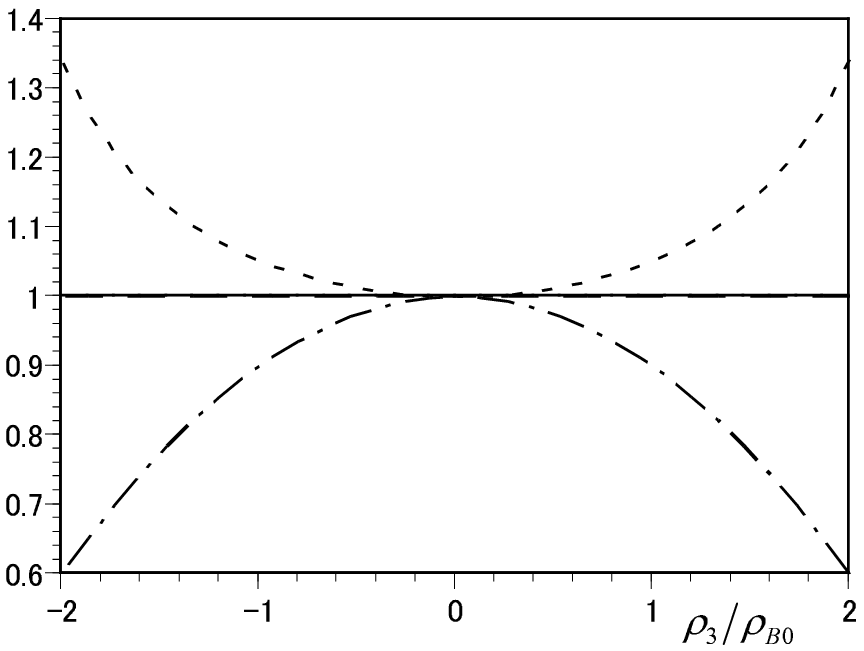}\\
fig.8
\end{tabular}
\end{center}

\begin{center}
\begin{tabular}{cc}
\includegraphics[height=70mm,width=90mm] {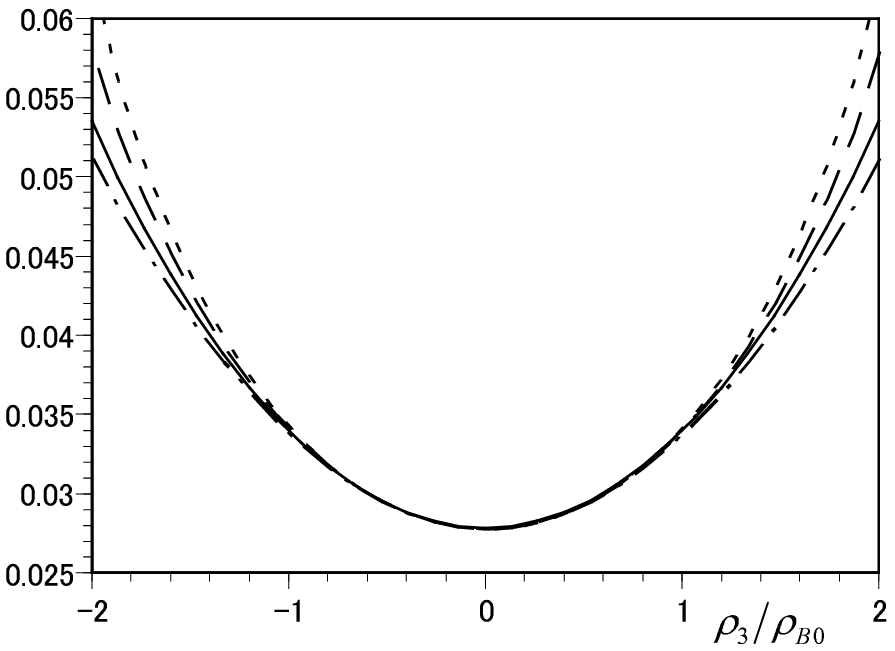}\\
fig.9
\end{tabular}
\end{center}

\newpage

\begin{center}
\begin{tabular}{cc}
\includegraphics[height=70mm,width=90mm] {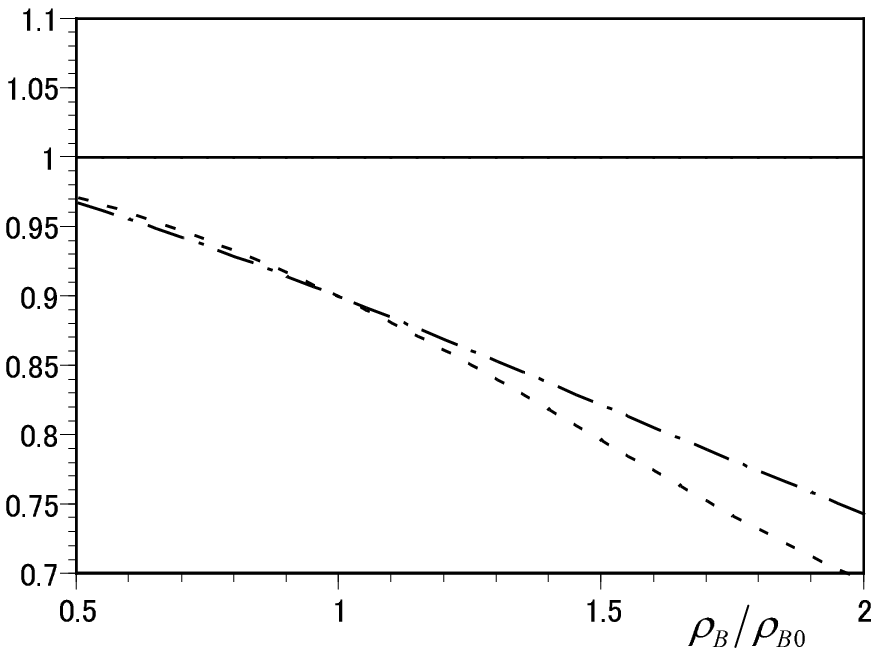} &
\includegraphics[height=70mm,width=90mm] {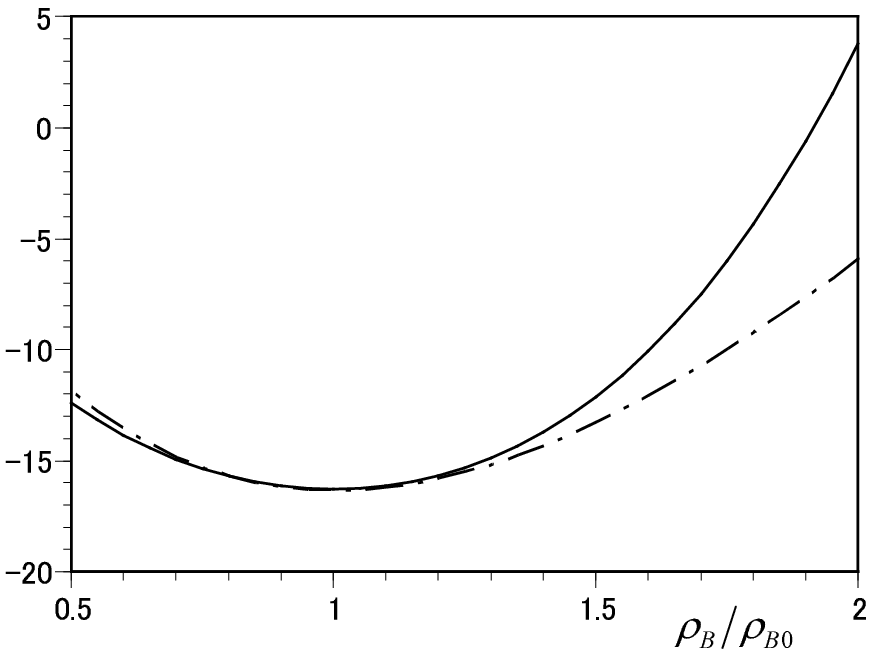}\\
fig.10
&
fig.11
\end{tabular}
\end{center}

\begin{center}
\begin{tabular}{cc}
\includegraphics[height=70mm,width=90mm] {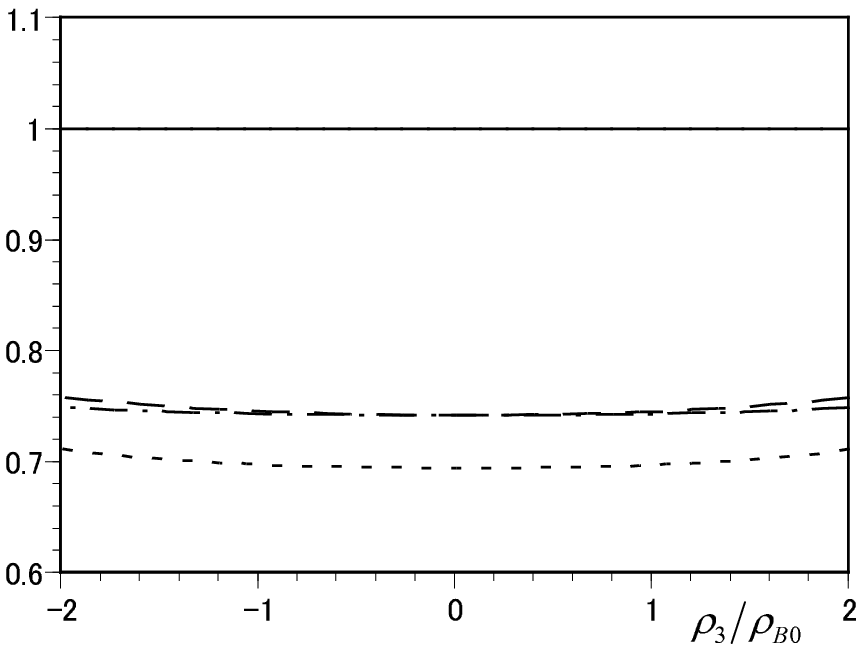} &
\includegraphics[height=70mm,width=90mm] {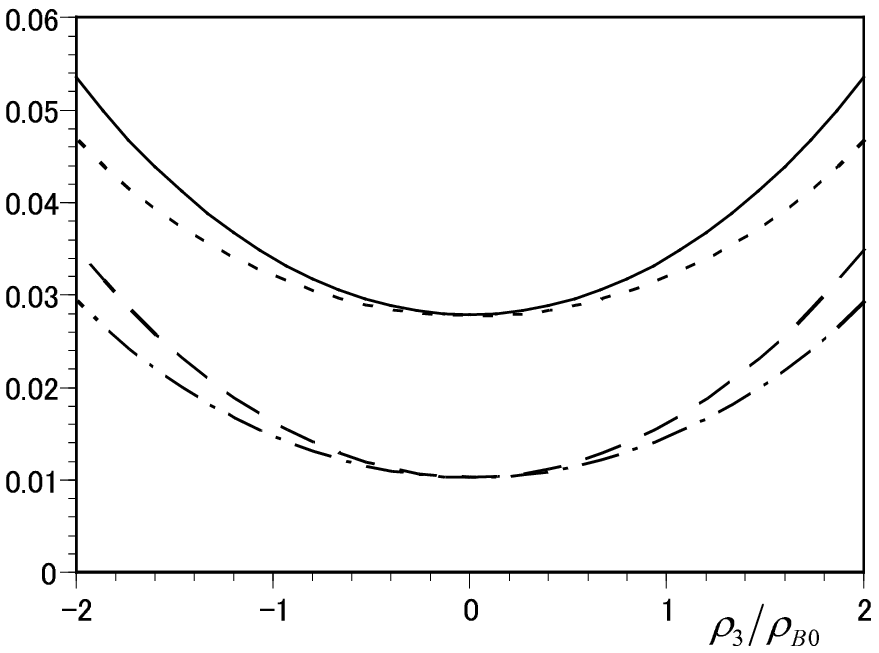}\\
fig.12
&
fig.13
\end{tabular}
\end{center}

\begin{center}
\begin{tabular}{cc}
\includegraphics[height=70mm,width=95mm] {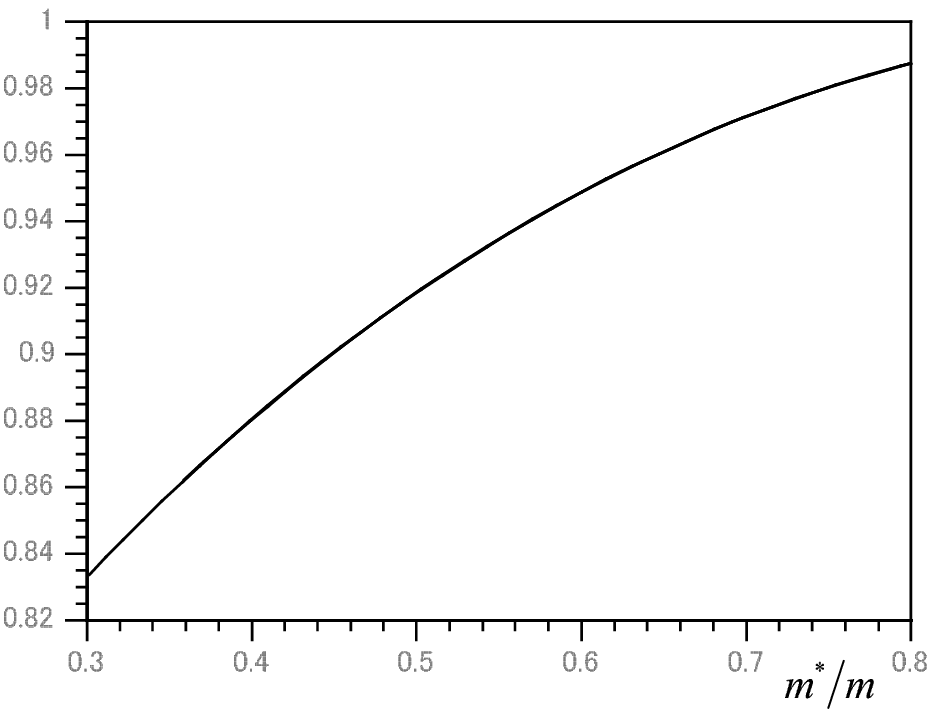} \\
fig.14
\end{tabular}
\end{center}

\begin{center}
\begin{tabular}{cc}
\includegraphics[height=70mm,width=90mm] {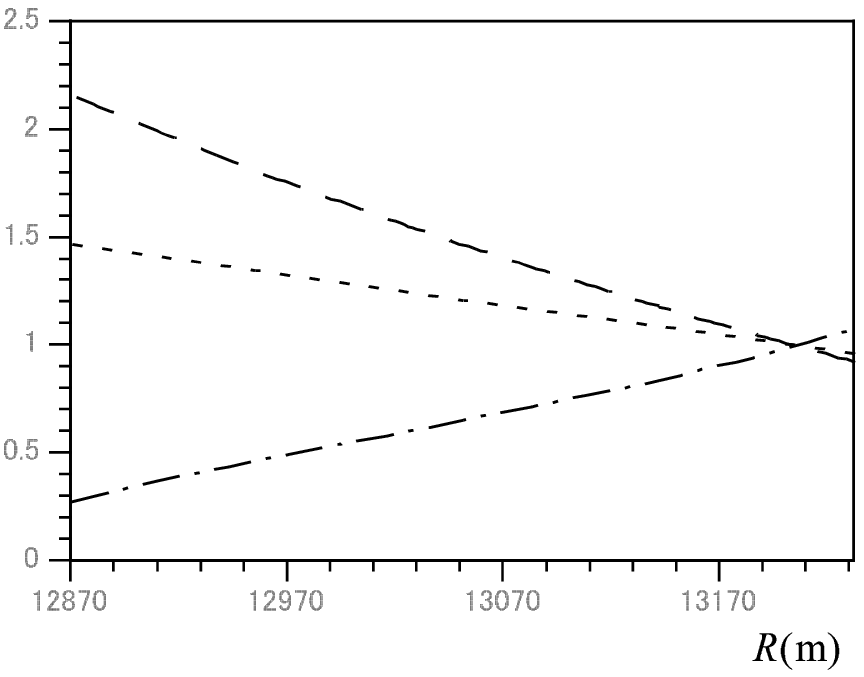}
\hspace{6mm} 
\includegraphics[height=70mm,width=95mm] {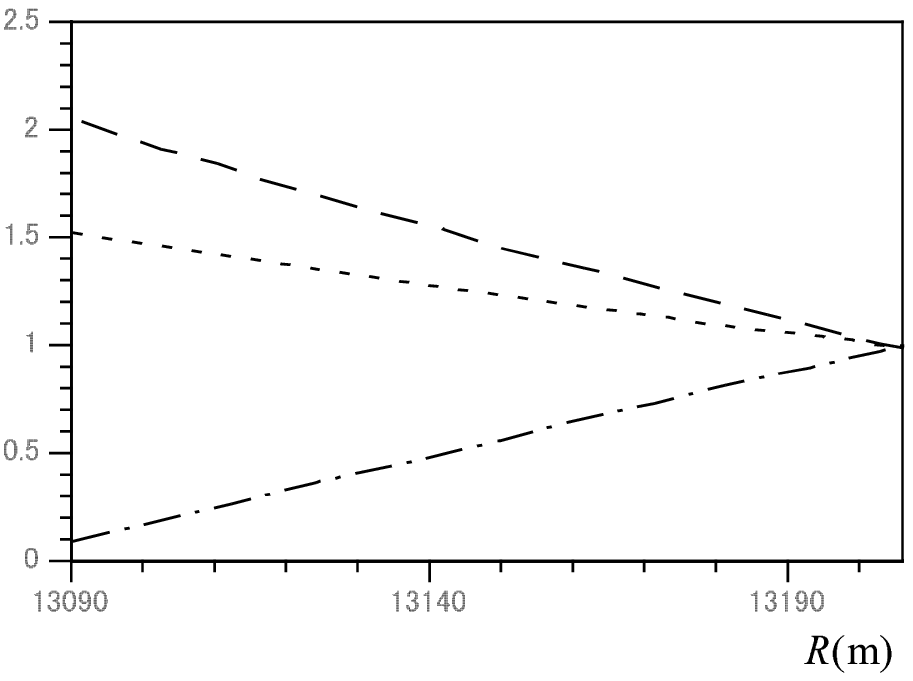}\\

fig.15 
\hspace{90mm}
fig.16
\end{tabular}
\end{center}

\begin{center}
\begin{tabular}{cc}
\includegraphics[height=70mm,width=92mm] {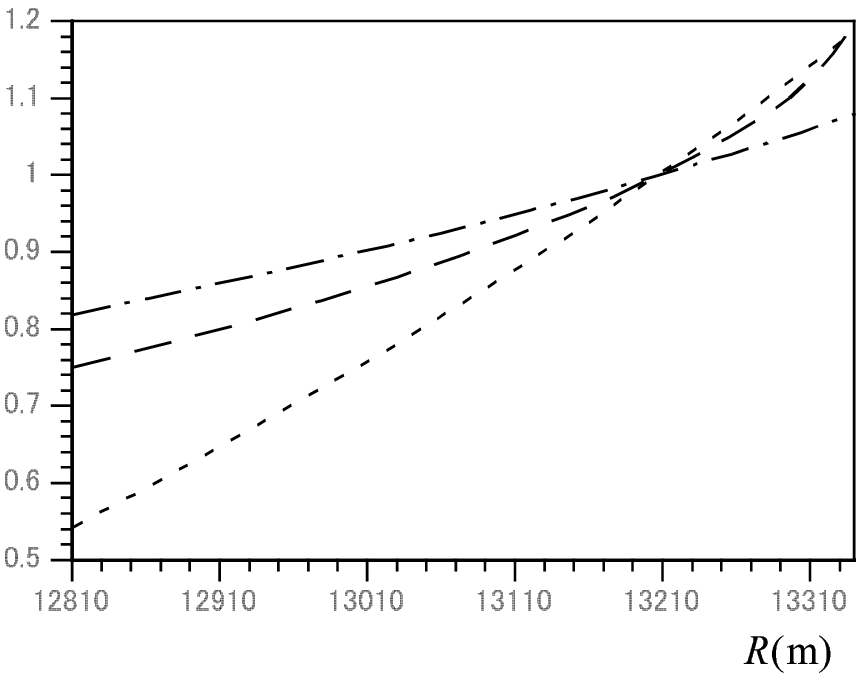}@
\hspace{1mm}
\includegraphics[height=70mm,width=91mm] {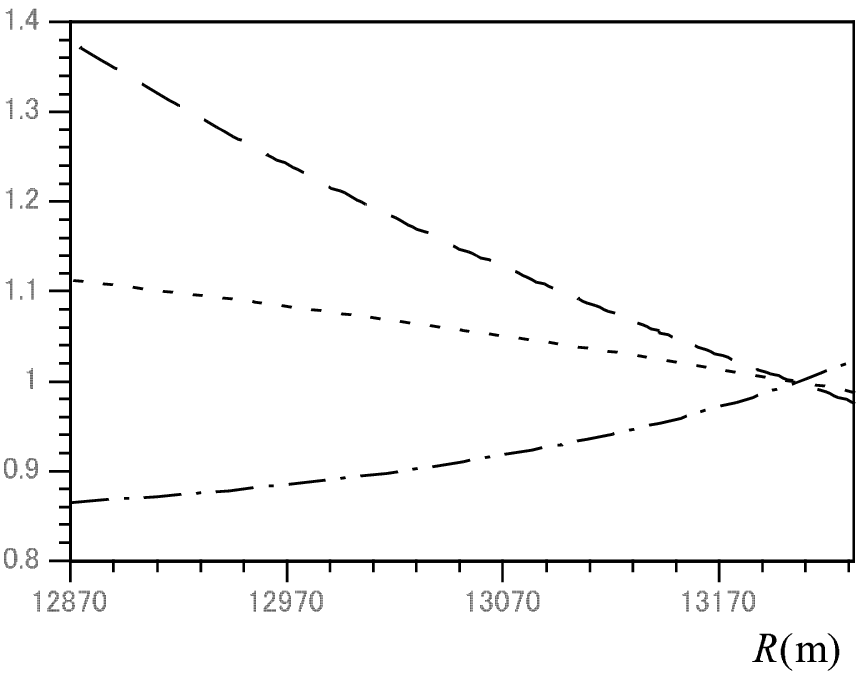} \\ 
fig.17
\hspace{90mm}
fig.18
\end{tabular}
\end{center}

\begin{center}
\begin{tabular}{cc}
\includegraphics[height=70mm,width=93mm] {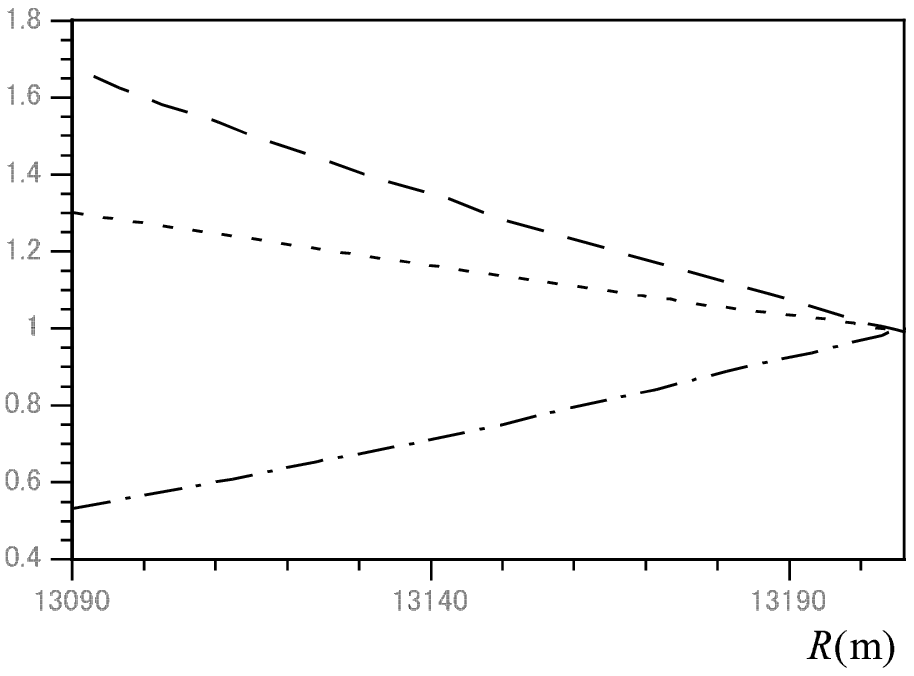} 
\hspace{2mm}
\includegraphics[height=70mm,width=93mm] {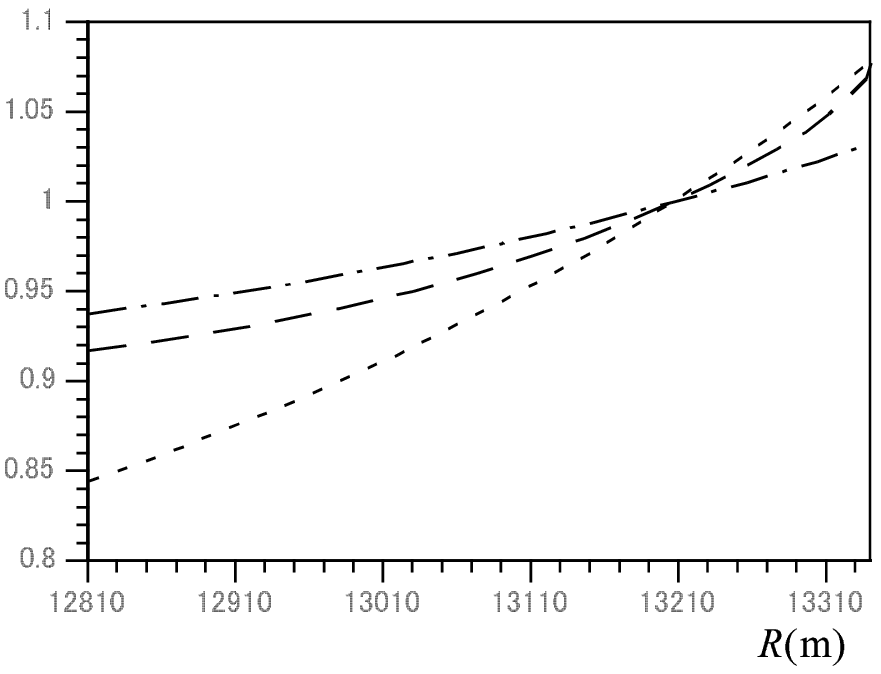}\\

fig.19 
\hspace{90mm}
fig.20

\end{tabular}
\end{center}

\newpage

\oddsidemargin -30mm

\begin{center}
\begin{tabular}{cc}
\includegraphics[height=70mm,width=91mm] {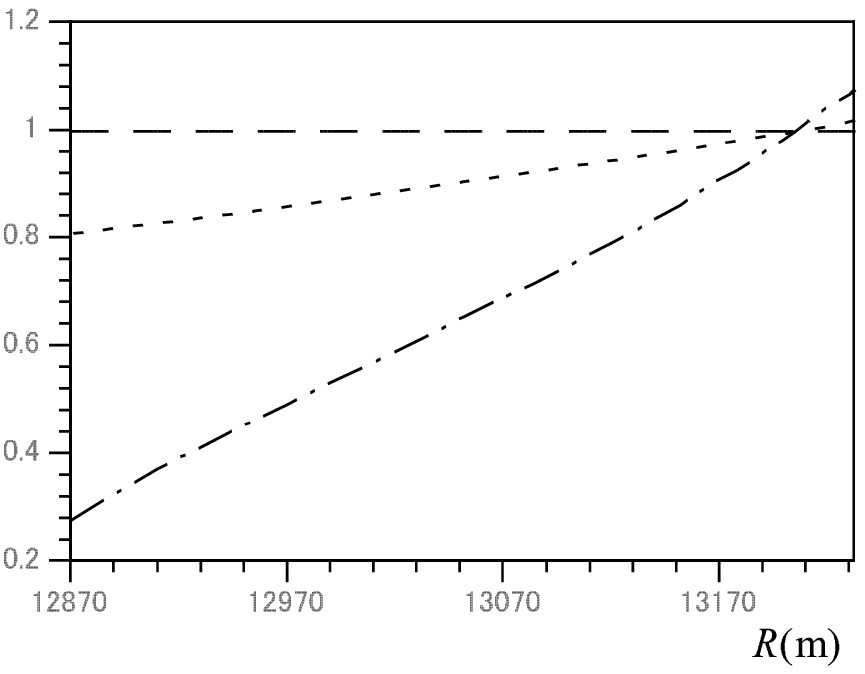}@
\includegraphics[height=70mm,width=90mm] {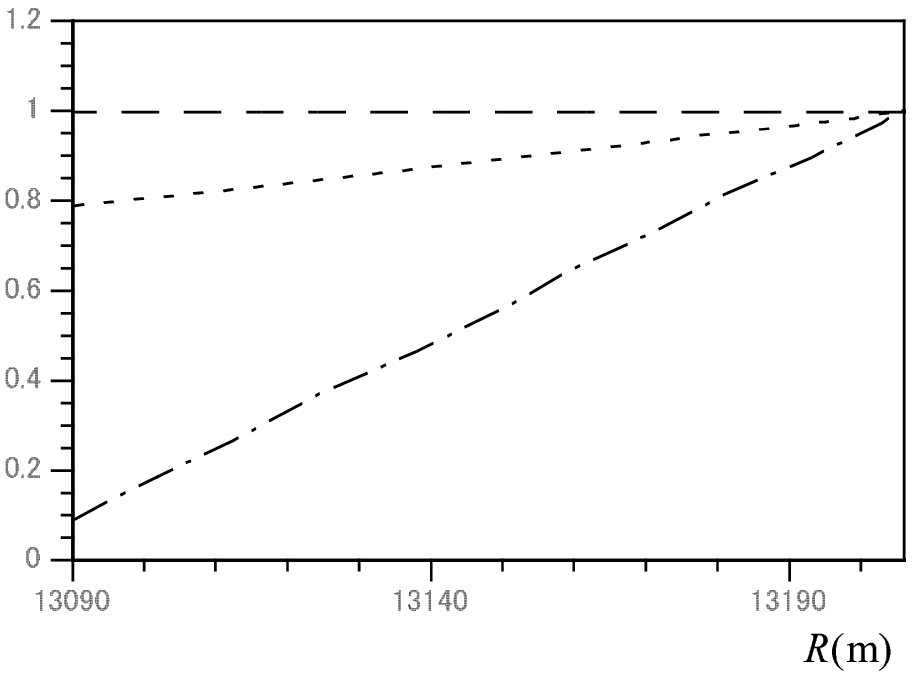} \\

fig.21
\hspace{90mm}
fig.22

\end{tabular}
\end{center}

\begin{center}
\begin{tabular}{cc}
\includegraphics[height=70mm,width=90mm] {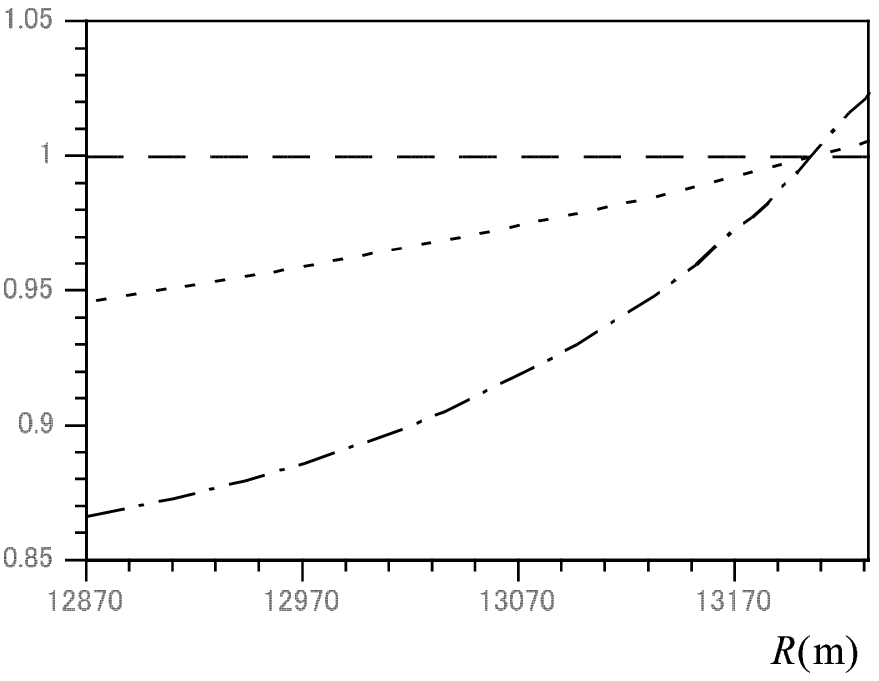} 
\hspace{2mm}
\includegraphics[height=70mm,width=90mm] {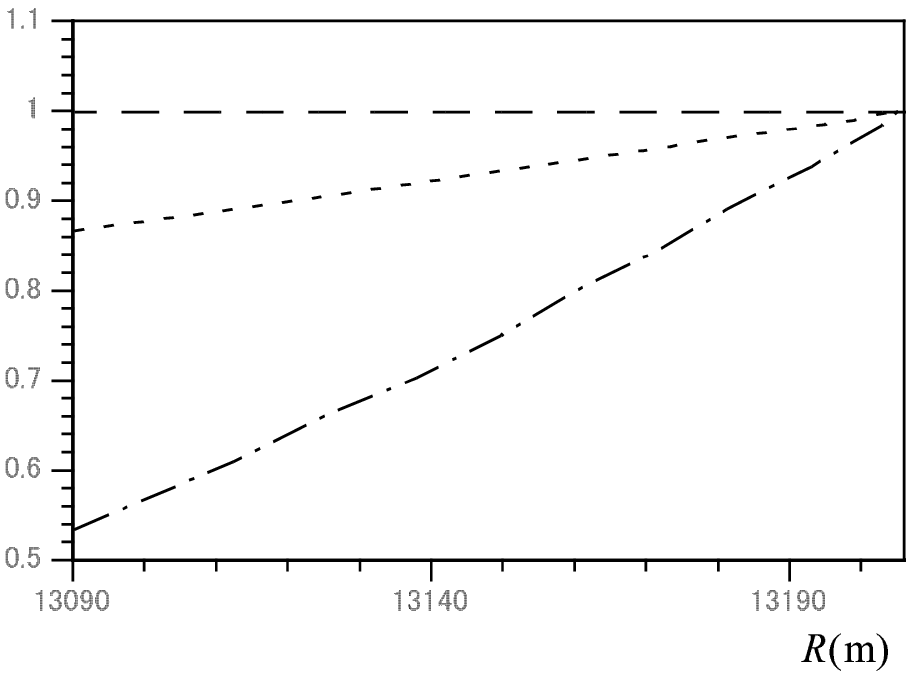}\\

fig.23 
\hspace{90mm}
fig.24
\end{tabular}
\end{center}

\end{center}

%\end{center}
%\end{center}

\end{document}